\documentclass{aa}  
\usepackage{adjustbox}

\usepackage{graphicx}
\usepackage{txfonts}

\usepackage[colorlinks=true,allcolors=blue]{hyperref}
\usepackage{color}
\usepackage{colortbl}
\usepackage{placeins}
\usepackage{orcidlink}
\usepackage{booktabs}
\usepackage{multirow}
\usepackage{longtable}
\usepackage{pdflscape}
\usepackage{array}
\usepackage{etoolbox}
\usepackage[toc,page]{appendix}

\begin{document} 

\title{Improving reddening estimates for RR Lyrae stars in the {\it Gaia} bands: a machine learning approach to the PAC(Z) relation}

                \author{A. Garofalo \orcidlink{0000-0002-5907-0375}
\inst{1}
          \and
          T. Muraveva\orcidlink{0000-0002-0969-1915}\inst{1}
          \and
          L. Monti \orcidlink{0000-0002-2087-0535}
          \inst{1}
          \and
          G. Clementini \orcidlink{0000-0001-9206-9723}
          \inst{1}
          \and
          F. Cusano \orcidlink{0000-0003-2910-6565}
          \inst{1}
          \and 
         M. L. Valentini \orcidlink{0009-0005-6787-5420}
          \inst{2}
}

  \institute{INAF - Osservatorio di Astrofisica e Scienza dello Spazio di Bologna, Via Gobetti 93/3, 40129 Bologna, Italy\\
              \email{alessia.garofalo@inaf.it}
       \and
Dipartimento di Fisica e Astronomia, Università di Bologna, Via Gobetti 93/2, 40129 Bologna, Italy
}

   \date{Received xxx; accepted xxx}

 
  \abstract
   {RR Lyrae stars are essential tracers of old stellar populations and distance indicators across the Milky Way (MW) and nearby galaxies. However, their use as standard candles is limited by uncertainties in extinction, particularly in the optical regime probed by the ESA mission {\it Gaia}. In {\it Gaia} Data Release 3 (DR3), individual absorption values (A$_G$) were provided for fundamental-mode RR Lyrae stars (RRab) based on an empirical period–amplitude–color (PAC) relation calibrated on a relatively small sample and on passband transformations.
   }
   {We aim to recalibrate extinction relations for RRab stars and, for the first time, establish analogous relations for first-overtone RR Lyrae stars (RRc) using {\it Gaia} DR3 photometry and pulsation parameters to provide intrinsic colors and interstellar absorption estimates.}
   {We used {\it Gaia} DR3 mean magnitudes and pulsation properties of RRab and RRc stars from an all-sky reference sample to recalibrate PAC(Z) relations in the {\it Gaia} passbands. Intrinsic colors $(G-G_{\rm RP})_0$ and $(G_{\rm BP}-G_{\rm RP})_0$ were derived through sequential feature selection combined with linear regression and bootstrap resampling, ensuring statistically consistent estimates of color excess and $A_G$.}
   {We recalibrated PACZ relations for RRab stars and derived, for the first time, PAC relations for RRc stars in the {\it Gaia} bands. The new relations reproduce intrinsic colors with residual scatters of $\sim$0.03–0.04 mag and provide $A_G$ estimates across both pulsation types. Tests on RR Lyrae stars in the MW field, globular clusters, and dwarf galaxy companions show consistent $A_G$ values, with the PAC(Z) relations based on $(G_{\rm BP}-G_{\rm RP})$ appearing more reliable than those using $(G-G_{\rm RP})$.}
   {The new relations provide consistent extinction estimates for both RRab and RRc stars, improving on the DR3 implementation and offering a valuable tool for Galactic studies and cosmic-distance scale work in view of {\it Gaia} DR4.}

   \keywords{Galaxy: halo --
                Galaxies: dwarf --
                Stars: variables: RR Lyrae 
               }
\titlerunning{{\it Gaia} DR3 RRLs Extinction relation(s) with ML techniques}
   \maketitle
%

\begin{figure*}
\center
\includegraphics[width=6.5cm]{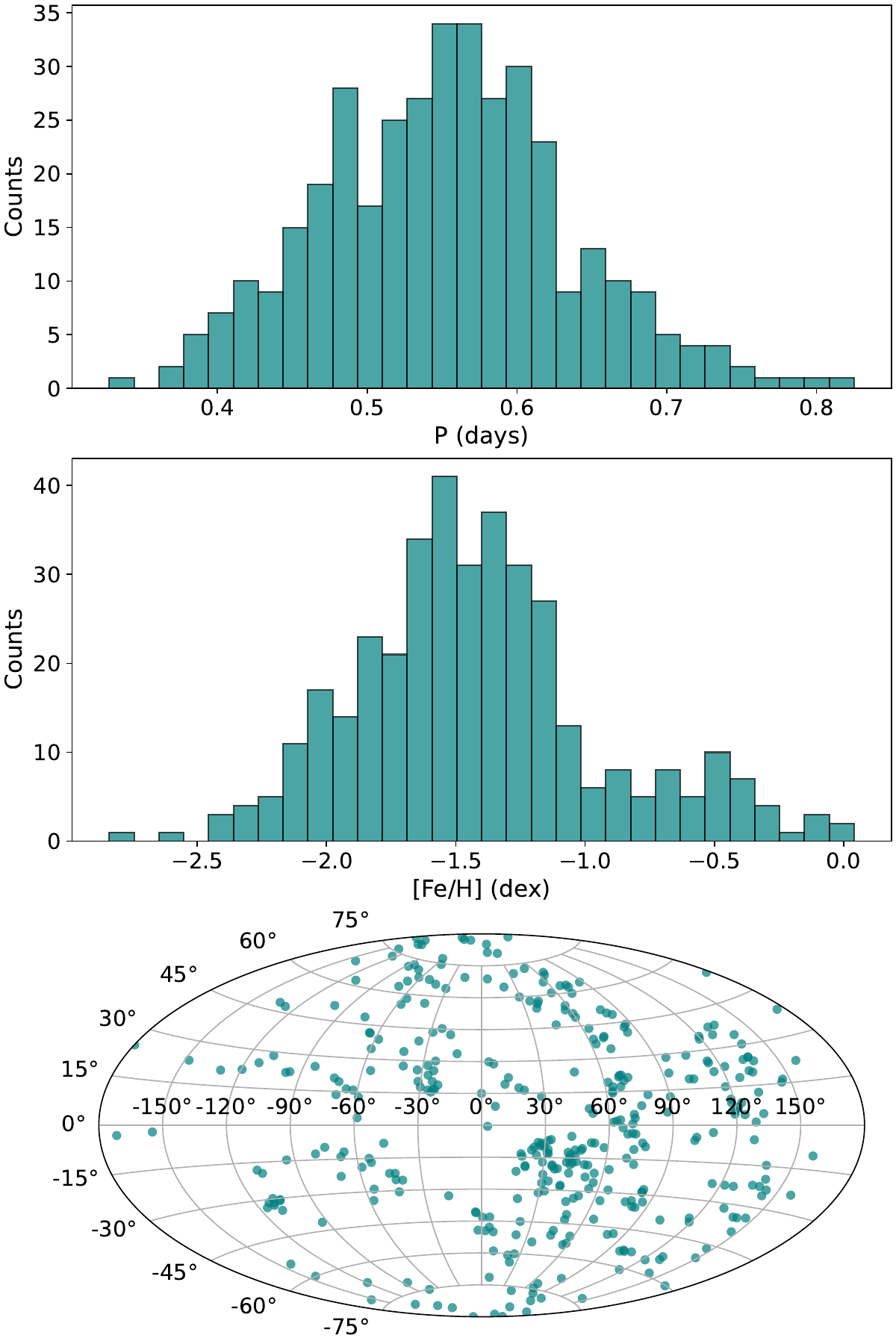}
~\includegraphics[width=6.5cm]{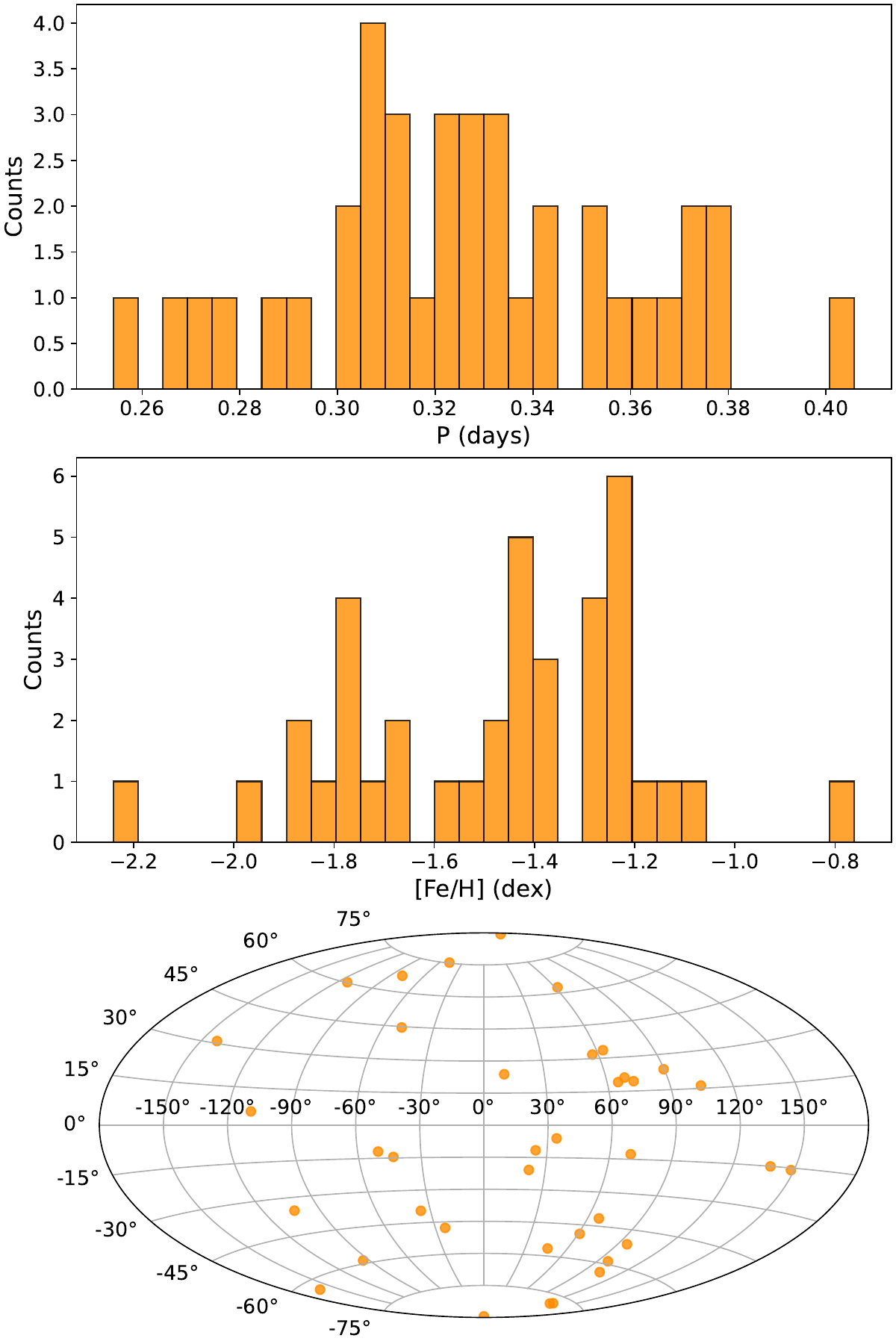}
    \caption{Distributions of pulsation periods (P) from the \textit{Gaia} DR3 \texttt{vari\_rrlyrae} table  (\citealt{Clementini-23}; top), metallicities ([Fe/H]) derived by \citet{Muhie-Dambis-2021}  in the \citet{ZW-1984} metallicity scale (middle), and sky position from the \textit{Gaia} DR3  \texttt{gaia\_source} table (\citealt{Gaiadr3-2023}; bottom) for the M21 final reference samples of RRab (373; tealblue) and RRc (38; orange) stars adopted in this study.}
    \label{fig:dt}
\end{figure*}
\section{Introduction}
RR Lyrae stars are the most numerous and best-studied Population II pulsating variables. 
They are easily identified thanks to their distinctive light-curve shapes, high amplitudes (up to 1.5–2 mag in the visual bands), and short pulsation periods (0.2–1 d), which require relatively modest temporal baselines for their characterization \citep[and references therein]{Catelan-2009,Catelan-2015}.\\
RR Lyrae  are valuable primary standard candles, following recognized luminosity–metallicity (LZ) relations in the  optical and period–luminosity–metallicity (PLZ) relations in the near-infrared \citep[e.g.][]{Bono-2003,Clementini-et-al-2003,Catelan-et-al-2004}. Equally important is their role as tracers of old stellar populations ($\gtrsim$10 Gyr). Given their broad sub-solar metallicity range ($-$2.5<[Fe/H]<0.2 dex), they are widely used to tracers of old (and typically metal-poor) stellar populations of the Milky Way (MW) and its satellites \citep{Smith-1995}. Due to their double role, RR Lyrae represent a fundamental tool for applications spanning from constraining stellar evolution to calibrating the Hubble constant through the cosmic distance ladder \citep[e.g.][and references therein]{Beaton-2016}.\\
Accurate distance determinations for RR Lyrae stars rely on well-calibrated LZ and PLZ relations, which are directly affected by uncertainties in parallax, metallicity, and extinction.
For decades, the limited availability and inhomogeneity of these quantities limited the accuracy of such relations. 
This situation has dramatically improved with the advent of all-sky surveys and space missions, which now provide unprecedented, homogeneous datasets of RR Lyrae stars. 
Among these surveys, the \emph{Gaia} mission has been transformative. 
Its third data release (DR3; \citealt{Gaiadr3-2023}) identifies almost 271,000 RR Lyrae stars distributed across the entire sky, tracing the MW halo, stellar streams, and the ancient components of MW satellite galaxies up to $G \simeq 21$ mag \citep{Clementini-23}.\\
For nearly all these stars, \emph{Gaia} DR3 provides astrometric (parallaxes, proper motions), photometric ($G$, $G_{BP}$, $G_{RP}$ mean magnitudes), and pulsation parameters (periods, amplitudes, Fourier coefficients, etc.).\\ 
In addition, \emph{Gaia} DR3 includes photometric metallicities for about 135,000 RR Lyrae stars and $G$-band absorption values ($A_G$) for roughly 140,000 fundamental-mode RR Lyrae (RRab) stars. 
These parameters were derived from empirical relations based on the pulsation characteristics 
measured directly from \emph{Gaia} $G$-band light curves of the RR Lyrae stars 
\citep{Clementini-2019, Clementini-23}.

Extinction is particularly relevant in the optical regime where \emph{Gaia} observes. 
Its spatially non-uniform distribution in the MW, especially toward the bulge and disk, represents one of the main sources of uncertainty in RR Lyrae-based distance estimates. 
At optical wavelengths, an uncertainty of $\sim0.2$ mag in $A_G$ can translate into distance errors of $\gtrsim 10\%$ for RR Lyrae at typical halo distances, 
whereas the impact can be even more relevant for sources in the Galactic bulge and disk, where extinction corrections represent a larger fraction of the total distance modulus.
\\
In \emph{Gaia} DR3, the $A_G$ values were obtained through a period–amplitude–color (PAC) relation involving the $G$-band amplitude, the pulsation period, and the $G - G_{RP}$ color \citep{Clementini-2019}. 
This relation was obtained by transforming to the Gaia photometric system the calibration of \citet{Piersimoni-2002} which was based on $V$ and $I$ photometry of about 60 RRab stars in the MW field  and two Globular clusters (GCs), IC 4499 and NGC 6362.\\
A limitation of the current $A_G$ estimates is that the underlying PAC relation is based on a small calibration sample, observed in different passbands from those of \emph{Gaia}.
Considering the intrinsic scatter of 0.02 mag in the \citet{Piersimoni-2002} relation and the uncertainties from photometric transformations \citep{Clementini-2019}, the global systematic error affecting \emph{Gaia} DR3 $A_G$ values for RRab stars is of the order of 0.22 mag. 
Typical total uncertainties, including observational and calibration errors, are around 0.46 mag. 
Minimizing such systematics will be crucial in the forthcoming \emph{Gaia} DR4 release which is based on 66 months of data and will provide improved light-curve sampling and more accurate pulsation parameters.\\
The advent of large and homogeneous datasets provided by \emph{Gaia} has accelerated the adoption of machine learning (ML) methods in stellar astrophysics, where they have proven particularly effective in uncovering empirical correlations among observables.\\
Recent works have exploited \emph{Gaia} RR Lyrae data to infer metallicities and other physical parameters from light-curve features. 
For instance, \citet{Muraveva-2025} used Sequential Feature Selection (SFS) to identify the most informative pulsation parameters for empirical metallicity relations, while \citet{Monti-2024, Monti-2025} applied Recurrent Neural Networks to estimate metallicities directly from \emph{Gaia} DR3 $G$-band light curves. 
These studies demonstrate the versatility and reliability of ML-based approaches in extracting astrophysical information from variability data.\\
In our work we employ the SFS algorithm \citep{Kohavi-1997,Guyon-2003} to explore the \emph{Gaia} DR3 parameter space and recalibrate the PAC relation for RRab stars. 
We also extend this approach to RRc stars, for which no analogous calibration currently exists in the literature.
SFS is a wrapper-based method that iteratively selects the subset of variables minimizing prediction error, providing robust and interpretable regression models while reducing overfitting.
Using the well-characterized Galactic RR Lyrae sample of \citet{Muhie-Dambis-2021} as a reference, we identify the parameters most strongly correlated with the intrinsic colors of RR Lyrae stars, leading to new empirical relations for both $(G - G_{RP})_0$ and $(G_{BP} - G_{RP})_0$.
These relations allow us to estimate individual color excesses, $E(G - G_{RP})$ and $E(G_{BP} - G_{RP})$, and thus derive $A_G$ values through established extinction coefficients. 
We test the validity of these relations across a variety of environments, including GCs, classical and ultra-faint dwarf galaxies, and the Magellanic Clouds (MCs). 
Independent extinction estimates available in the literature are used for external validation, providing a robust assessment of the applicability of our new PAC relations across different astrophysical contexts.\\
This paper is organized as follows.
Section~\ref{sec:data} describes the selection and preparation of the Galactic RR Lyrae dataset used as a calibration reference.
Section~\ref{sec:feature_selection} outlines the feature–selection procedure and validation of the regression models used to derive intrinsic colors and extinction estimates.
Section~\ref{sec:ris} presents the recalibrated PAC(Z) relations for RRab and RRc stars and compares the resulting $A_G$ values across different environments — from the all-sky field sample to GCs, dwarf galaxies, and the MCs.
Finally, our conclusions are summarized in Section~\ref{sec:con}.

\section{Galactic RR Lyrae sample selection}\label{sec:data}
%
\begin{table}
	\centering
    \footnotesize
     \caption{Main properties of our final reference samples M21 (373 RRab and 38 RRc) selected from \citet{Muhie-Dambis-2021}.}
	\label{tab:dt}
	\begin{tabular}{lcccc} 
		\hline
  Sample&  N & P  & [Fe/H]& $\langle G \rangle$\\
        &    & (d)& (dex) & (mag)\\

  \hline
  M21~RRab &373&[0.33,0.82]&[$-$2.84,0.04]&[8.93,17.68]\\
  M21~RRc &38&[0.25,0.40]&[$-$2.24,$-$0.76]&[9.01,14.51]\\
		\hline
	\end{tabular}
\end{table}

To derive new PAC(Z) relations for RR Lyrae stars in the \emph{Gaia} bands, for RRab and RRc stars separately, we combined 
sample selection with ML techniques.\\
We selected a dataset of RR Lyrae stars located in different Galactic environments and characterized by independently determined physical properties (e.g. metallicity, extinction). This provided a reliable reference set for training and validating the empirical relations
allowing us to estimate intrinsic color indices, $(G - G_{\rm RP})_0$ and $(G_{\rm BP} - G_{\rm RP})_0$, from which we inferred the color excesses $E(G - G_{\rm RP})$, $E(G_{BP} - G_{RP})$ and the A$_{G}$ values (see Section~\ref{sec:ris}). 
The RRab and RRc datasets adopted as reference to train and validate our models are those published by \citet{Muhie-Dambis-2021}.
 They include 850 well-characterised field RR Lyrae stars (751 RRab and 99 RRc stars) with photometric and spectroscopic data, cross-identified with {\it Gaia} DR2. In particular, the broad coverage in pulsation period (from 0.219 to 0.956 days) and  metallicity (from $-$2.84 to 0.07 dex) makes this dataset a robust reference sample for feature selection and ML training.\\
Metallicities for 448 of the 850 RR Lyrae stars in \citet{Muhie-Dambis-2021} sample were derived by these authors from a new homogeneous spectroscopic dataset obtained with
the Robert Stobie Spectrograph on the Southern African Large Telescope (SALT).
For the remaining part of the sample, metallicities were compiled from earlier literature sources \citep{Dambis-2013}. 
 All these values were subsequently homogenised onto the \citet{ZW-1984} scale by \citet{Muhie-Dambis-2021}, ensuring a consistent metallicity framework across the entire reference dataset despite the variety of the original metallicity estimates. Pulsation periods and modes for 
about half of the stars were also taken from 
\citet{Dambis-2013}, which consolidated periods from ASAS3, the \citet{Maintz-2005} catalogue, and the General Catalogue of Variable Stars \citep{Samus-2017}. For the remaining objects, periods and modes were newly derived through time-series analysis of dedicated  South African Astronomical Observatory photometry combined with All-Sky Automated Survey for Supernovae light curves, ensuring phase coherence with the SALT spectroscopic observations.\\
\begin{figure}
\center
~\includegraphics[width=\columnwidth]{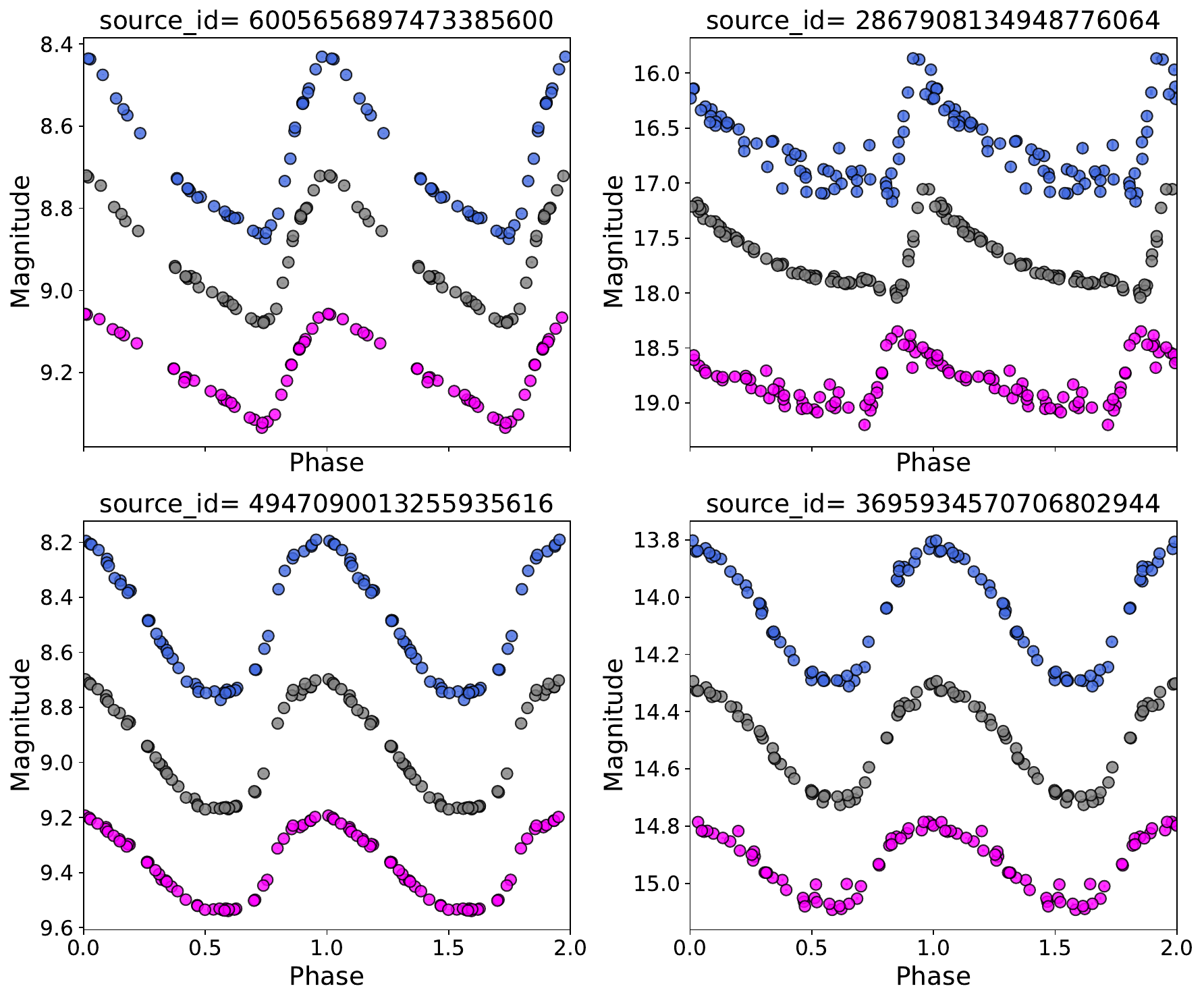}
    \caption{{\it Gaia} DR3 phase-folded light curves for four RR Lyrae stars in the $G$ (grey), $G_{\rm BP}$ (blue), and $G_{\rm RP}$ (magenta) bands, selected from the M21 subsamples. The curves are repeated over two pulsation cycles (phase $\in [0, 2]$) for clarity. 
    The $G_{\rm BP}$ and $G_{\rm RP}$ magnitudes have been vertically shifted by $-0.5$ and $+0.7$ mag, respectively, to enhance visual separation.
    The upper panels show two RRab stars, while the lower panels show two RRc stars. 
    Photometric uncertainties 
    are generally smaller than 
    the symbol size. The phase-folding was performed using the {\it Gaia} DR3 periods, which are in excellent agreement with values reported in the literature.} 
    \label{fig:dt_lc}
\end{figure}

We cross-matched these sources with the {\it Gaia} DR3 clean catalogue of 270,891 RR Lyrae variables \citep[\texttt{vari\_rrlyrae} table;][]{Clementini-23}. The matched stars were further filtered to retain only those satisfying two  criteria:
 \begin{itemize}
 \item{\texttt{num\_clean\_epochs\_g}$\geq$ 40 } 
 \item{$\Delta(P_{literature}-P_{Gaia~DR3}$)$\leq$ 0.001 d}
  \end{itemize}
These criteria ensure that the mean magnitudes, pulsation periods, amplitudes, and Fourier parameters published in {\it Gaia} DR3, and used in this study, are derived from well-sampled and reliable light curves.\\
In particular, the first condition guarantees that all selected RR Lyrae stars have sufficient epoch coverage to confidently exclude the presence of undetected double-mode pulsators (RRd stars), since the search for secondary periodicities in {\it Gaia} DR3 was only performed for sources with at least 40 clean $G$-band epochs (see \citealt{Clementini-23}, section~2). The second condition ensures that the pulsation periods adopted from DR3 are fully consistent with the literature values. This, in turn, provides confidence that the associated mean magnitudes, amplitudes, and Fourier parameters, either published or derived from DR3 light curves, are reliable and suitable for subsequent analyses.\\
After applying these criteria, our final reference samples include 373 RRab and 38 RRc stars. These are our reference RR Lyrae samples which are hereafter referred to as M21. Their main properties are summarised in Table~\ref{tab:dt} and Figure~\ref{fig:dt}, which show that the M21 sources are distributed all-sky for both RRab and RRc stars. 
\\ Given the predominantly all-sky distribution of the M21 samples, we adopted a standard extinction law with \( R_V = 3.1 \) for all sources. This choice ensures consistency in the extinction correction and reflects the fact that most of our sources are not concentrated in strongly reddened regions. 
We acknowledge that in some areas of the sky, in particular toward the Galactic bulge, the extinction law may deviate from the standard one, often showing lower values of \( R_V \) \citep[and references therein]{Nataf-2013}. However, since only a small fraction of our sources lie in those regions, adopting a uniform extinction law provides a good compromise between accuracy and homogeneity.\\
\citet{Muhie-Dambis-2021} derived the absorption in the $V$ band (\(A_{V}\)) by applying the 3D extinction model of \citet{Drimmel-2003} and implementing the iterative approach outlined by \citet{Dambis-2013}.
  For stars located at low Galactic latitudes (\(|b| \le 40^\circ\))
,where the 3D map becomes unreliable due to the clumpy dust distribution, \(A_{V}\) values were instead derived from color excesses \(E(V - W1)\) or \(E(V - K_s)\), calibrated through period–metallicity–color relations and converted using the extinction coefficients of \citet{Yuan-2013}. \\
We first converted the M21 $A_{V}$ values into $E(B-V)$ to derive the individual absorptions $A_{G}$, $A_{G_{BP}}$, and $A_{G_{RP}}$. For this purpose, we adopted the total-to-selective extinction ratios $R_{G} = 2.516 \pm 0.036$, $R_{G_{BP}} = 3.266 \pm 0.037$, and $R_{G_{RP}} = 1.936 \pm 0.025$ from \citet{Huang-2021}.  
The intrinsic colors $(G - G_{\rm RP})_0$ and $(G_{\rm BP} - G_{\rm RP})_0$ were then obtained by correcting the \texttt{int\_average\_g}, \texttt{int\_average\_bp}, and \texttt{int\_average\_rp} magnitudes from \textit{Gaia} DR3 \texttt{vari\_rrlyrae} table \citep{Clementini-23} for the corresponding absorptions, i.e. $A_{G}$, $A_{G_{BP}}$, and $A_{G_{RP}}$.\\
To build the reference sample used to calibrate the empirical relations for intrinsic colors, we adopted the \texttt{int\_average\_g}, \texttt{int\_average\_bp}, and \texttt{int\_average\_rp} magnitudes from the {\it Gaia} DR3 \texttt{vari\_rrlyrae} table \citep{Clementini-23}. 
These parameters are obtained through model-based integration of the light curves, previously cleared of major outliers, which better captures the pulsational variability of RR Lyrae stars compared to the simple flux averages (\texttt{phot\_mean\_mag}) from the \texttt{gaia\_source} table \citep{Gaiadr3-2023}. In the case of our reference sample, composed of relatively bright stars with high signal-to-noise and well-sampled light curves, the difference between \texttt{int\_average} and \texttt{phot\_mean} magnitudes is negligible, but the former provides a more physically consistent description of the variability.\\
When extending the analysis to the full all-sky RR Lyrae sample (Section~\ref{sec:ris}), we adopted a mixed approach: we used \texttt{int\_average\_g}, which is always well defined, and \texttt{phot\_bp\_mean\_mag} and \texttt{phot\_rp\_mean\_mag}. The latter are more robust and less sensitive to modeling uncertainties than their \texttt{int\_average} counterparts in the low signal-to-noise regime, which typically affects the $G_{BP}$ and $G_{RP}$ light curves, particularly for fainter sources.\\
Figure~\ref{fig:dt_lc} shows the light curves of the brightest (left) and faintest (right) RRab (top) and RRc (bottom) stars included in our calibration data set. 
These examples show that the \texttt{int\_average} magnitudes are based on well-sampled light curves, providing reliable mean magnitudes across both pulsation modes and magnitude ranges.\\
The resulting reference samples and extinction-corrected photometry provide the basis for the feature selection and regression modelling described in Sect.~\ref{sec:feature_selection}.
 The full parameter sets for the M21 calibration sample are detailed in Tables~\ref{tab:rrlab_params_full} and \ref{tab:rrlc_params_full} of Appendix~\ref{sec:appendix_a}.

\begin{table}[h]
\centering
\small 
\caption{Summary of the features tested in the regression models to predict $(G - G_{\rm RP})_0$ and $(G_{\rm BP} - G_{\rm RP})_0$ intrinsic colors.}
\label{tab:features}
\begin{tabular}{p{1.8cm} p{6.5cm}}
\toprule
Feature & Description \\
\midrule
$\log P$ & Logarithm of the pulsation period \\
$P$ & Pulsation period for RRab and RRc stars \\
$\phi_{21, G}$ & Fourier phase difference $\phi_2 - 2\phi_1$ (second-to-first harmonic, $G$ band) \\
$\phi_{31, G}$ & Fourier phase difference $\phi_3 - 3\phi_1$ (third-to-first harmonic, $G$ band) \\
$r_{21, G}$ & Fourier amplitude ratio $A_2/A_1$ (second-to-first harmonic, $G$ band) \\
$r_{31, G}$ & Fourier amplitude ratio $A_3/A_1$ (third-to-first harmonic, $G$ band) \\
$\mathrm{Amp}G$ & \texttt{peak\_to\_peak\_g}, amplitude in $G$ band \\
$\mathrm{Amp}G_{\mathrm{BP}}$ & \texttt{peak\_to\_peak\_bp}, amplitude in $G_{\mathrm{BP}}$ band \\
$\mathrm{Amp}G_{\mathrm{RP}}$ & \texttt{peak\_to\_peak\_rp}, amplitude in $G_{\mathrm{RP}}$ band \\
$\mathrm{[Fe/H]}$ & Metallicity in \citet{ZW-1984} scale \\ 
\bottomrule
\end{tabular}
\end{table}

\begin{figure*}[h!]
\centering
\includegraphics[width=16cm]{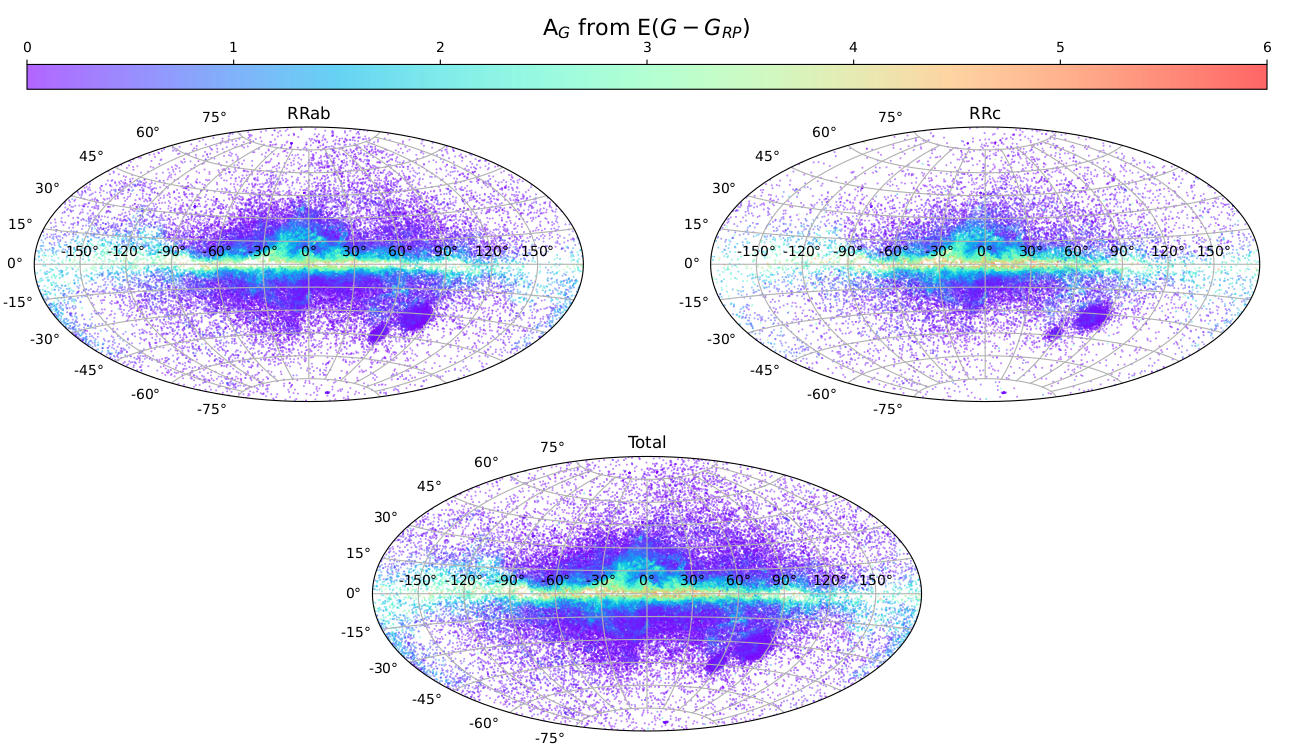}
\includegraphics[width=16cm]{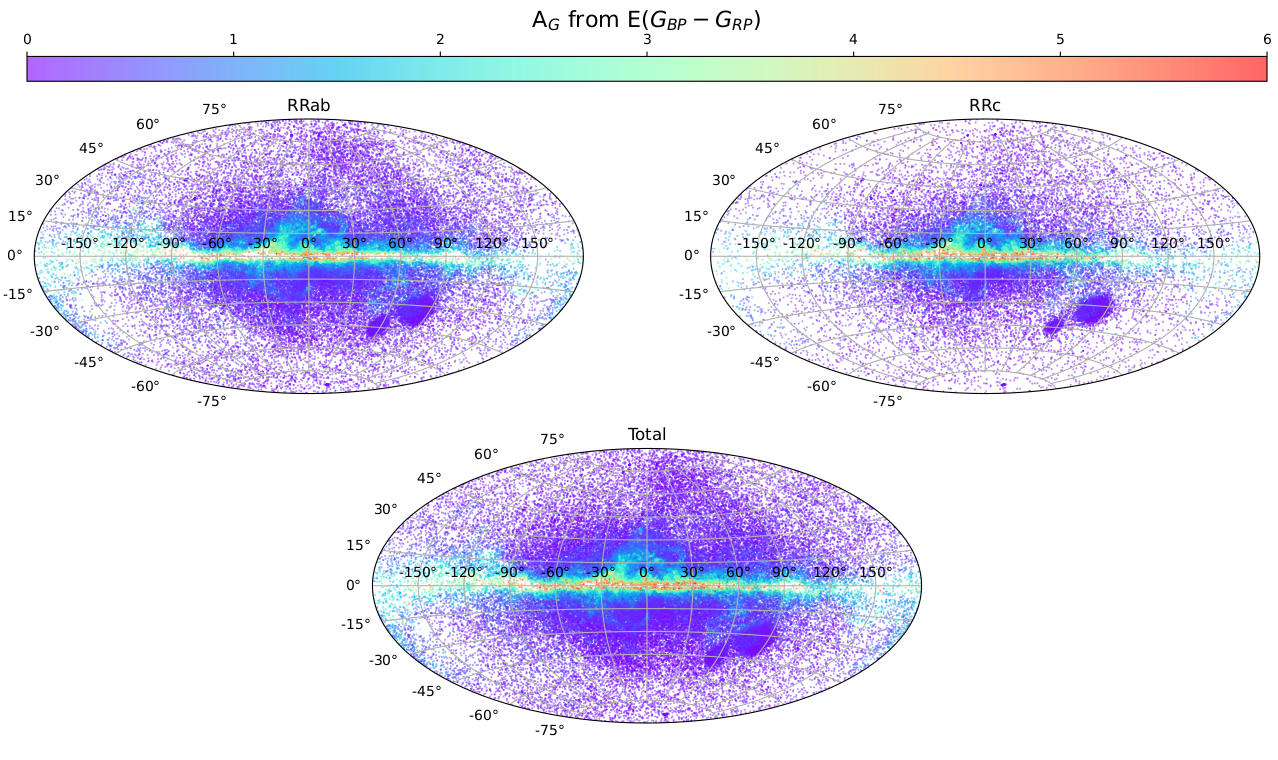}
\caption{{\it Top panels}: All-sky distribution in Galactic coordinates of the absorption in the $G$ band ($A_G$) for the whole clean sample of RR Lyrae stars in the {\it Gaia} DR3 \texttt{vari\_rrlyrae} table. {\it Top panels:} $A_G$ derived from $E(G - G_{\mathrm{RP}})$ using the relations reported in Table~\ref{tab:regression_formulas}, for RRab stars (top left, $N_{\rm RRab} = 90802$, with $A_{G}\ge 0$ 70308), RRc stars (top right, $N_{\rm RRc} = 51238$, with $A_{G}\ge 0$ 42276, using AmpG), and the combined RRab+RRc sample (bottom, $N_{\rm RRab+RRc} = 142040$). {\it Bottom panels:} Same as in the top panels but for $A_G$ derived from $E(G_{\mathrm{BP}} - G_{\mathrm{RP}})$, with $N_{\rm RRab} = 89915$ (with $A_{G}\ge 0$ 87745), $N_{\rm RRc} = 51238$ (with $ A_{G}\ge 0$ 49178), and $N_{\rm RRab+RRc} = 141153$. The colorbars range from 0 to 6 mag; individual negative values are not included in the plots for physical visualization but are fully retained in the statistical analysis.}  
      \label{fig:allsky_ag_bprpcu5}
\end{figure*}
\section{Feature selection and model validation}\label{sec:feature_selection}
We describe here the procedure used to identify the most informative features for predicting the intrinsic colors of RR Lyrae stars, and the approach adopted to validate the resulting regression models, ensuring robust and unbiased estimates. Specifically, Section~\ref{sec:features_sfs} details the input features and the setup of the SFS algorithm, while Section~\ref{sec:results_sfs} presents the results of the feature selection and the corresponding regression analysis.
\begin{figure*}
\sidecaption
  \begin{minipage}{12cm}
    \centering
    \includegraphics[width=12cm]{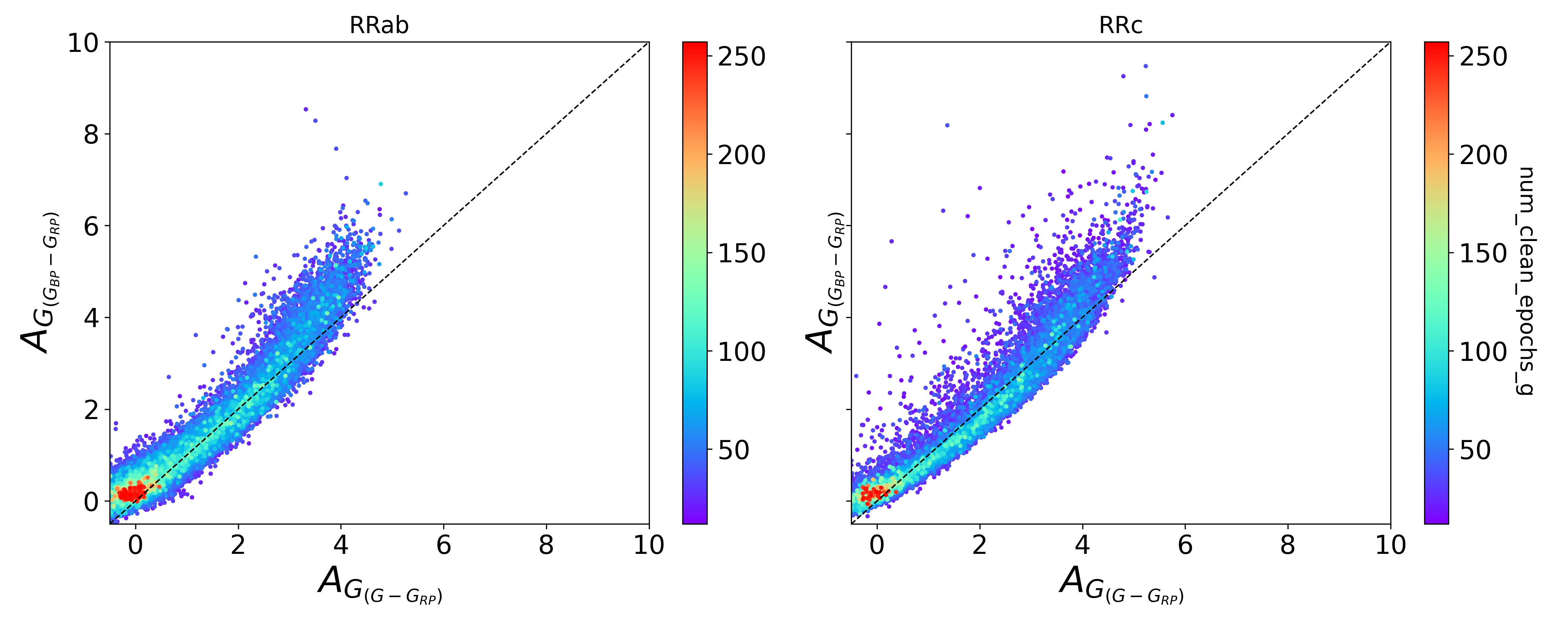} \\[0.1cm] 
    \includegraphics[width=12cm]{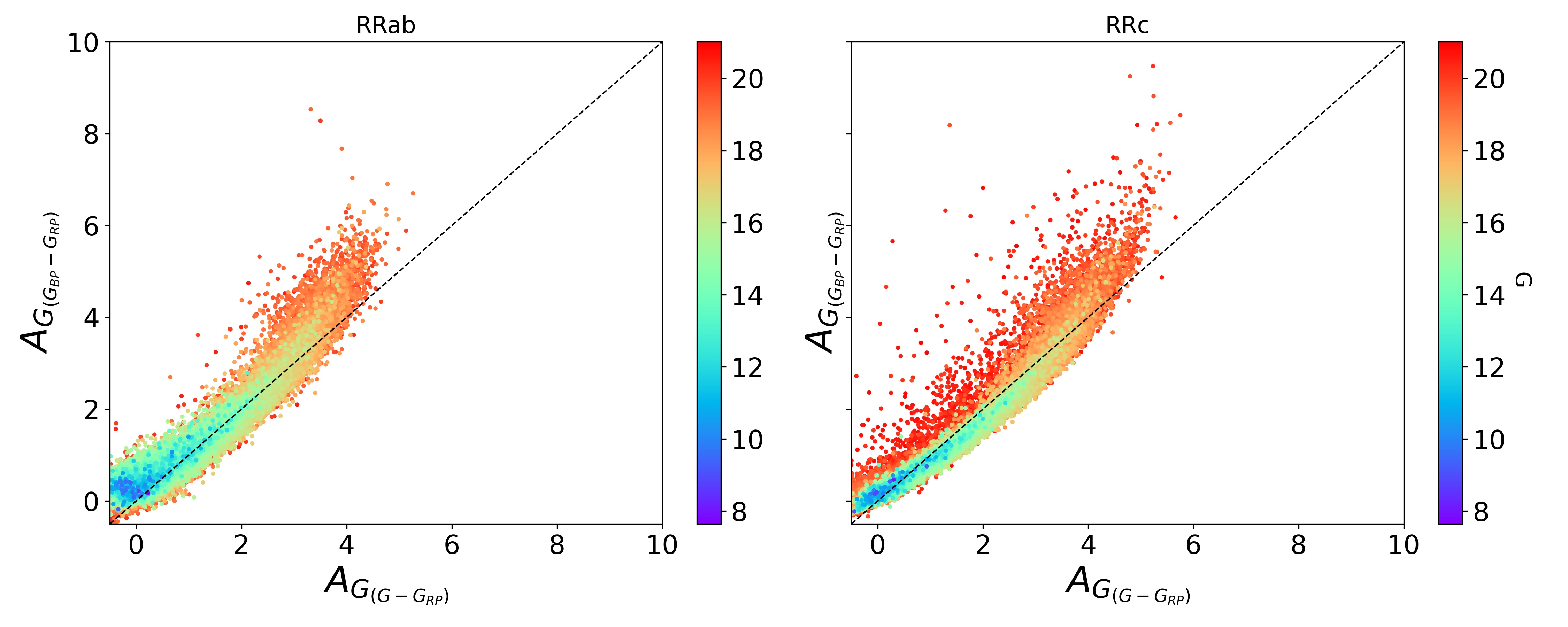}
  \end{minipage}
  \caption{Comparison between $A_G$ values derived from the $(G_{\rm BP}-G_{\rm RP})$ relation and the $(G-G_{\rm RP})$ relation for the RRab (left) and RRc (right) samples. Black, dashed lines represent 1:1 relations. Points are color-coded according to the number of $G$-band epochs (\texttt{num\_clean\_epochs\_g}; upper panels) and by the intensity-averaged $G$-band magnitude (lower panels). Brighter sources are plotted in the foreground to better show their distribution.}
  \label{fig:gc_confronto_colori}
\end{figure*}
\subsection{Input features and setup of the SFS procedure}\label{sec:features_sfs}
The same SFS algorithm and model validation procedure were applied separately to RRab and RRc stars. 
The SFS algorithm was coupled with a linear regression model to determine the combination of pulsation and metallicity parameters that best predict the intrinsic colors.
The set of features tested in the regression analysis for both subclasses is summarised in Table~\ref{tab:features}. These include:
\begin{itemize}
    \item Pulsation periods, Fourier parameters ($\phi_{21, G}$, $\phi_{31, G}$, $r_{21, G}$, $r_{31, G}$), and peak-to-peak amplitudes in the \textit{Gaia} bands (AmpG, AmpG${_{BP}}$, and AmpG${_{RP}}$), from the \textit{Gaia} DR3 \texttt{vari\_rrlyrae} table \citep{Clementini-23};
    \item Metallicity estimates on the \citet{ZW-1984} scale for our M21 reference samples from \citep{Muhie-Dambis-2021}.
\end{itemize}

These quantities formed the initial pool of candidate {\it predictors} for the SFS procedure, which identified the parameters that best reproduce the target variables of our models: the intrinsic {\it Gaia} colors $(G_{\rm BP}-G_{\rm RP})_0$ and $(G-G_{\rm RP})_0$. These targets were computed using {\it Gaia} DR3 \texttt{int\_average} magnitudes, corrected for individual absorption according to the procedure in Section~\ref{sec:data}.\\
The algorithm was applied independently to RRab and RRc stars, given their distinct pulsational properties and accuracies. Using a forward-selection approach with linear regression as the base estimator, features were iteratively added according to the improvement in mean squared error (MSE), evaluated through 5-fold cross-validation. The 5-fold configuration was chosen as an optimal balance between bias and variance for our sample size, ensuring stable estimates without excessively fragmenting the data.\\
At each iteration, the feature whose addition most improves the performance of the model is included in the selected subset. We systematically explored models ranging from one to all available features, recording the cross-validated MSE and its standard deviation at each step.\\
This approach allowed us to determine the optimal number of features that balance predictive performance and model simplicity. 
The use of cross-validation was crucial in mitigating the risk of overfitting, particularly given the relatively small sample size of RR Lyrae datasets. 
This approach identifies the most informative parameters and preserves interpretability, allowing direct comparison with classical PAC(Z) relations. The selected features and validation procedure were then applied to derive the empirical relations, as described in the following section.

\begin{table*}
\caption{Linear regression models for $(G - G_{\rm RP})_0$ and $(G_{\rm BP} - G_{\rm RP})_0$ intrinsic colors of RRab and RRc stars, in our M21 
samples. The coefficients are derived using SFS-selected features and uncertainties correspond to the standard deviation from bootstrap resampling.}
\label{tab:regression_formulas}
\centering
\tiny
\begin{tabular}{ll}
\toprule
Relation & RMS\\
\midrule
   RRab&\\
\hline
\\
 $(G - G_{\rm RP})_0=-0.068(\pm 0.008)\times\mathrm{AmpG} + 0.393(\pm 0.034)\times \mathrm{logP} + 0.030(\pm 0.004)\times\mathrm{[Fe/H]} + 0.548 (\pm 0.011)$&0.027\\
\\
  $(G_{\rm BP} - G_{\rm RP})_0 =  - 0.087(\pm 0.011) \times\mathrm{Amp{G_{BP}}} + 0.820(\pm 0.055)\times \mathrm{logP} + 0.051(\pm 0.007)\times\mathrm{[Fe/H]} + 0.874(\pm 0.017) $& 0.044\\
\midrule
 RRc&\\
\hline
\\
 $(G - G_{\rm RP})_0 = -  0.139 (\pm 0.069) \times \mathrm{AmpG_{BP}} + 0.365 (\pm 0.123) \times \mathrm{logP} + 0.518 (\pm 0.078)$ &0.029\\ 
 $(G - G_{\rm RP})_0 = -  0.164 (\pm 0.104) \times \mathrm{AmpG} + 0.383 (\pm 0.130) \times \mathrm{logP} + 0.526 (\pm 0.092)$ &0.029\\
\\
 {\bf $(G_{\rm BP} - G_{\rm RP})_0 = -0.295 (\pm 0.131) \times \mathrm{AmpG} + 0.808(\pm 0.102) \times \mathrm{logP}+0.896(\pm 0.102)$}&0.037\\
\bottomrule
\end{tabular}
\end{table*}

\subsection{Results of feature selection and regression analysis}\label{sec:results_sfs}
The results of this procedure are summarized in Figure~\ref {fig:elbow_ab} and~\ref {fig:elbow_c} in Appendix~\ref{sec:appendix_a}, which show the evolution of the MSE as a function of the number of selected features.
For the RRab sample, the prediction error, MSE, shows a clear plateau after the inclusion of the first 3 features (see Figure~\ref{fig:elbow_ab}), indicating that additional parameters provide only marginal improvement. This behaviour suggests that the intrinsic colors of RRab stars can be reliably predicted from a compact set of features (logP, AmpG or AmpG$_{BP}$ and [Fe/H]), with limited risk of overfitting.  The MSE values at each SFS step, corresponding to the incremental addition of one new feature at a time, are listed in Table~\ref{tab:mse_ab}. 
In particular, the features identified as most informative by the SFS procedure correspond to those that define the classical PACZ relations already established in the literature. This confirms that our data-driven approach successfully recovers the physically relevant predictors without introducing spurious correlations, and underscores the robustness of the methodology.\\
In contrast, for the RRc sample the prediction error begins to increase when adding more features beyond the first two, as shown in both panels of Figure~\ref{fig:elbow_c}. This trend indicates that the additional parameters do not enhance predictive power of the model and may instead introduce noise. The corresponding MSE values at each step of the SFS procedure are listed in Table~\ref{tab:mse_c}.
The absence of a clear plateau may reflect both the smaller sample size and the weaker dependence of the intrinsic colors on secondary parameters for this subclass.
Therefore, for RRc stars, we retained only the first two most informative features to construct the regression model for intrinsic color predictions.
We note that for $(G-G_{\rm RP})_0$, the two informative feature combinations identified by the SFS are the period $P$ together with either the AmpG or Amp$G_{BP}$. 
Both combinations provide essentially identical residuals. 
However, since AmpG is more precisely measured and available for a larger fraction of {\it Gaia} DR3 RRc stars, in the analysis and figures presented in this paper we adopt the $P$--AmpG relation for estimating $(G-G_{\rm RP})_0$. 
For completeness, both relations are reported in Table~\ref{tab:regression_formulas}.
This result highlights a fundamental difference between the two types: while RRab stars conform to a well-defined PACZ-like behaviour, RRc stars follow a simpler, more compact predictive scheme.\\
To quantify the uncertainties in the regression coefficients and the intercept obtained from the SFS, we used a bootstrap resampling approach. 
We performed 1000 bootstrap resamplings to estimate the uncertainties on the regression coefficients for each sample. 
The standard deviation of the bootstrapped coefficients was used as an empirical measure of uncertainty.
This approach allows us to provide robust estimates of the regression parameters along with their associated errors, accounting for sample variability without assuming parametric error distributions. The final coefficients and their uncertainties are reported in Table~\ref{tab:regression_formulas}. 
\\
Figures~\ref{fig:lr_m21_ab} and~\ref{fig:lr_m21_c} (in Appendix~\ref{sec:appendix_a}) show the comparison between predicted and observed intrinsic colors, along with the root mean square (RMS) of the residuals, which provides a quantitative measure of the model accuracy.

\begin{figure}[h]
\center
~\includegraphics[width=\columnwidth]{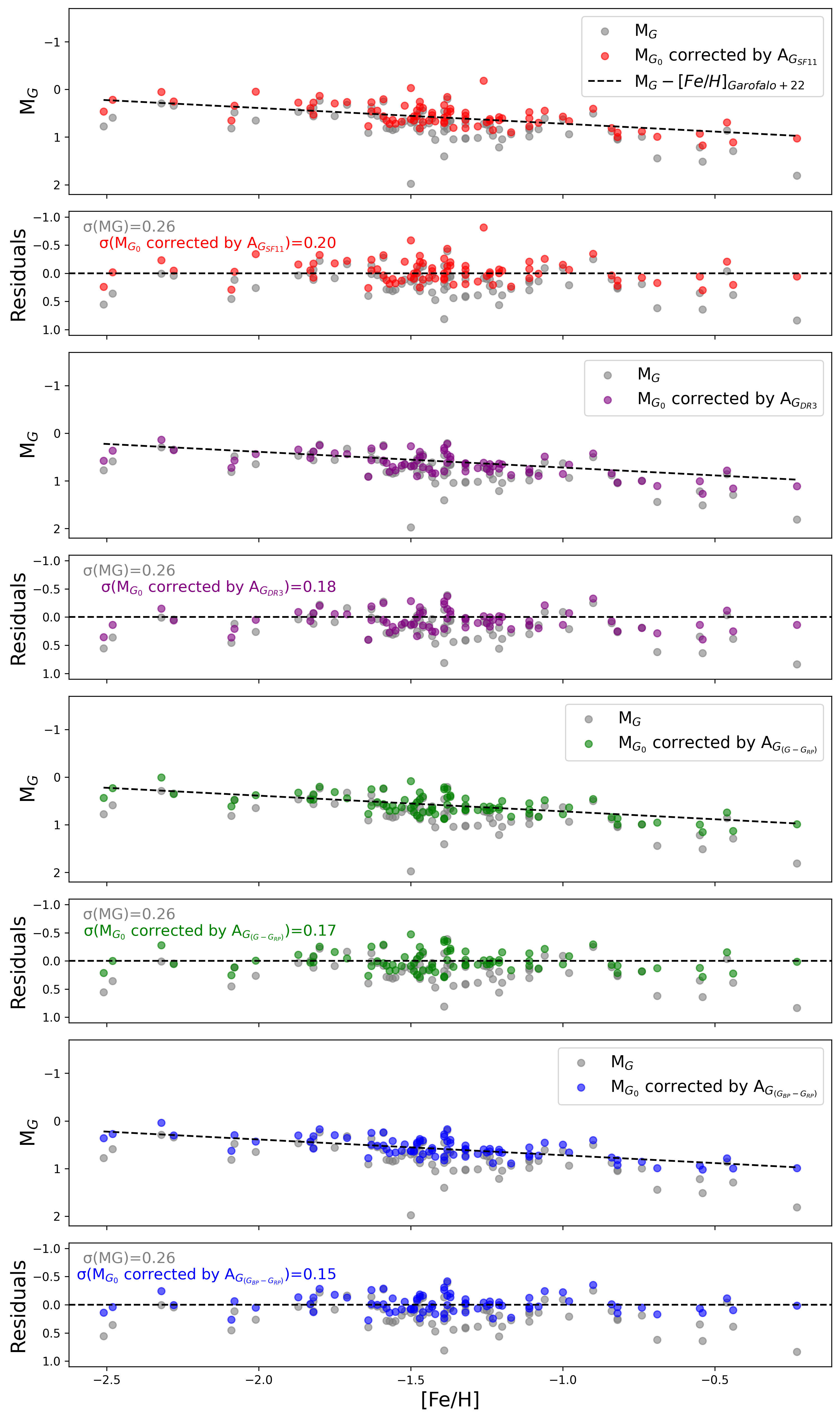}
    \caption{Comparison of $G$-band absolute magnitudes for 80 Galactic field RR Lyrae stars from \citet{Garofalo-et-al-2022} as a function of metallicity. Stars in common with the M21 training set were excluded. The main panels show {\it Gaia} DR3 
    M$_{G}$ (gray, uncorrected) and absorption-corrected (M$_{G_{0}}$) using A$_{G_{SF11}}$ (red; SF11)
    , A$_{G_{DR3}}$ (purple; \citealt{Clementini-23}), and our A$_{G}$ based on ($G-G_{\rm RP}$; green) and ($G_{\rm BP}-G_{\rm RP}$; blue).  Magnitudes are derived from sources with positive parallaxes and relative uncertainties $<20\%$. The bottom sub-panels show the residuals with respect to the  M$_{G}$--[Fe/H] relation from \citet[][eq. 19]{Garofalo-et-al-2022}.}
    \label{fig:lz}
\end{figure}

\section{Results and comparisons}\label{sec:ris}

\begin{table*}
\tiny
\centering
\caption{
Mean values and standard deviation of the E($G-G_{RP}$), E($G_{BP}-G_{RP}$) color excesses, and the corresponding $A_G$ values for RR Lyrae stars in the selected dSphs (classical and ultra-faint).}
\label{tab:all_sistemi}
\begin{tabular}{llcclcclclc}
\toprule
System & Type & E($G-G_{RP}$) & A$_{G_{(G-G_{RP})}}$ & N & E($G_{BP}-G_{RP}$) & A$_{G_{(G_{BP}-G_{RP})}}$ & N & A$_{G_{DR3}}$ & N \\
& & (mag) & (mag) & & 
(mag) & (mag) &  & (mag) & \\
\midrule
Carina & ab & $-$0.042 $\pm$ 0.227 & $-$0.181 $\pm$ 0.984 & 4 & 0.025 $\pm$ 0.229 & 0.047 $\pm$ 0.433 & 4 & 0.913 $\pm$ 1.174 & 15 \\
Carina & c & 0.007 $\pm$ 0.103 & 0.029 $\pm$ 0.449 & 3 & 0.153 $\pm$ 0.148 & 0.289 $\pm$ 0.280 & 3 & $-$ & $-$ \\
Carina & ab+c & $-$0.021 $\pm$ 0.173 & $-$0.091 $\pm$ 0.751 & 7 & 0.080 $\pm$ 0.196 & 0.150 $\pm$ 0.370 & 7 & $-$ & $-$ \\
\midrule
Coma & ab & 0.037 $\pm$ 0.044 & 0.161 $\pm$ 0.190 & 1 & 0.102 $\pm$ 0.082 & 0.193 $\pm$ 0.154 & 1 & 0.052 $\pm$ 0.098 & 1 \\
Coma & c & $-$0.031 $\pm$ 0.042 & $-$0.133 $\pm$ 0.184 & 1 & $-$0.059 $\pm$ 0.061 & $-$0.112 $\pm$ 0.111 & 1 & $-$ & $-$ \\
Coma & ab+c & 0.003 $\pm$ 0.048 & 0.014 $\pm$ 0.208 & 2 & 0.022 $\pm$ 0.114 & 0.041 $\pm$ 0.216 & 2 & $-$ & $-$ \\
\midrule
Draco & ab & $-$0.022 $\pm$ 0.073 & $-$0.095 $\pm$ 0.315 & 80 & 0.079 $\pm$ 0.109 & 0.149 $\pm$ 0.207 & 79 & 0.435 $\pm$ 0.356 & 150 \\
Draco & c & $-$0.039 $\pm$ 0.082 & $-$0.169 $\pm$ 0.357 & 32 & 0.042 $\pm$ 0.119 & 0.080 $\pm$ 0.225 & 32 & $-$ & $-$ \\
Draco & ab+c & $-$0.027 $\pm$ 0.076 & $-$0.116 $\pm$ 0.328 & 112 & 0.068 $\pm$ 0.113 & 0.129 $\pm$ 0.213 & 111 & $-$ & $-$ \\
\midrule
Hydrus~I & ab & 0.096 $\pm$ 0.045 & 0.416 $\pm$ 0.194 & 2 & 0.151 $\pm$ 0.043 & 0.286 $\pm$ 0.081 & 2 & 0.245 $\pm$ 0.083 & 4 \\
Hydrus~I & c & $-$ & $-$ & $-$ & $-$ & $-$ & $-$ & $-$ & $-$ \\
Hydrus~I & ab+c & 0.096 $\pm$ 0.045 & 0.416 $\pm$ 0.194 & 2 & 0.151 $\pm$ 0.043 & 0.286 $\pm$ 0.081 & 2 & $-$ & $-$ \\
\midrule
Sextans & ab & $-$0.010 $\pm$ 0.090 & $-$0.045 $\pm$ 0.390 & 9 & 0.145 $\pm$ 0.197 & 0.273 $\pm$ 0.372 & 9 & 0.896 $\pm$ 1.594 & 40 \\
Sextans & c & 0.008 $\pm$ 0.128 & 0.036 $\pm$ 0.554 & 18 & 0.116 $\pm$ 0.215 & 0.220 $\pm$ 0.406 & 18 & $-$ & $-$ \\
Sextans & ab+c & 0.002 $\pm$ 0.115 & 0.009 $\pm$ 0.499 & 27 & 0.126 $\pm$ 0.206 & 0.237 $\pm$ 0.389 & 27 & $-$ & $-$ \\
\midrule
Tucana~II & ab & $-$0.029 $\pm$ 0.039 & $-$0.127 $\pm$ 0.122 & 1 & 0.125 $\pm$ 0.065 & 0.236 $\pm$ 0.122 & 1 & 0.291 $\pm$ 0.140 & 1 \\
Tucana~II & c & $-$0.099 $\pm$ 0.023 & $-$0.431 $\pm$ 0.098 & 2 & $-$0.012 $\pm$ 0.056 & $-$0.023 $\pm$ 0.106 & 2 & $-$ & $-$ \\
Tucana~II & ab+c & $-$0.076 $\pm$ 0.044 & $-$0.330 $\pm$ 0.189 & 3 & 0.034 $\pm$ 0.089 & 0.063 $\pm$ 0.168 & 3 & $-$ & $-$ \\
\midrule
Tucana~III & ab & 0.001 $\pm$ 0.031 & 0.006 $\pm$ 0.134 & 3 & 0.085 $\pm$ 0.081 & 0.160 $\pm$ 0.154 & 3 & 0.049 $\pm$ 0.143 & 4 \\
Tucana~III & c &  $-$0.047 $\pm$ 0.019 &  $-$0.206 $\pm$ 0.084 & 1 &  $-$0.004 $\pm$ 0.031 & $-$0.008 $\pm$  0.055 & 1 & $-$ & $-$ \\
Tucana~III & ab+c &  $-$0.011 $\pm$ 0.035 & $-$0.047 $\pm$ 0.152 & 4 & 0.062 $\pm$ 0.080 & 0.118 $\pm$ 0.151 & 4 & $-$ & $-$ \\
\midrule
UMa~I & ab &  $-$0.048 $\pm$ 0.051 &  $-$0.208 $\pm$ 0.222 & 3 & 0.033 $\pm$ 0.091 & 0.062 $\pm$ 0.173 & 3 & 2.744 $\pm$ 2.754 & 3 \\
UMa~I & c &  $-$ &  $-$ &  $-$ &  $-$ &  $-$ &  $-$ &  $-$ &  $-$ \\
UMa~I & ab+c &  $-$0.048 $\pm$ 0.051 &  $-$0.208 $\pm$ 0.222 & 3 & 0.033 $\pm$ 0.091 & 0.062 $\pm$ 0.173 & 3 &  $-$ &  $-$ \\
\midrule
UMa~II & ab & 0.067 $\pm$ 0.009 & 0.291 $\pm$ 0.039 & 2 & 0.208 $\pm$ 0.002 & 0.392 $\pm$ 0.004 & 2 & 0.263 $\pm$ 0.052 & 2 \\
UMa~II & c & 0.112 $\pm$ 0.064 & 0.488 $\pm$ 0.277 & 2 & 0.272 $\pm$ 0.138 & 0.514 $\pm$ 0.260 & 2 & $-$  & $-$ \\
UMa~II & ab+c & 0.090 $\pm$ 0.045 & 0.390 $\pm$ 0.197 & 4 & 0.240 $\pm$ 0.088 & 0.453 $\pm$ 0.166 & 4 & $-$  & $-$  \\
\midrule
Bootes~I & ab & $-$0.049 $\pm$ 0.051 & $-$0.212 $\pm$ 0.223 & 3 & 0.013 $\pm$ 0.059 & 0.025 $\pm$ 0.112 & 3 & 0.073 $\pm$ 0.095 & 4 \\
Bootes~I & c & $-$0.001 $\pm$ 0.046 & $-$0.003 $\pm$ 0.198 & 7 & 0.076 $\pm$ 0.057 & 0.143 $\pm$ 0.108 & 7 & $-$  & $-$ \\
Bootes~I & ab+c & $-$0.015 $\pm$ 0.050 & $-$0.066 $\pm$ 0.218 & 10 & 0.057 $\pm$ 0.062 & 0.108 $\pm$ 0.118 & 10 & $-$  & $-$  \\
\midrule
Bootes~II & ab & $-$0.034 $\pm$ 0.050 & $-$0.149 $\pm$ 0.216 & 1 & 0.044 $\pm$ 0.072 & 0.083 $\pm$ 0.136 & 1 & $-$0.074 $\pm$ 0.124 & 1 \\
Bootes~II & c & $-$  & $-$  & $-$  & $-$  & $-$  & $-$  & $-$  & $-$  \\
Bootes~II & ab+c & $-$0.034 $\pm$ 0.050 & $-$0.149 $\pm$ 0.216 & 1 & 0.044 $\pm$ 0.072 & 0.083 $\pm$ 0.136 & 1 & $-$  & $-$  \\
\midrule
Bootes~III & ab & $-$0.071 $\pm$ 0.058 & $-$0.310 $\pm$ 0.252 & 2 & 0.029 $\pm$ 0.022 & 0.055 $\pm$ 0.042 & 2 & 0.008 $\pm$ 0.117 & 1 \\
Bootes~III & c & $-$0.033 $\pm$ 0.034 & $-$0.143 $\pm$ 0.147 & 3 & 0.033 $\pm$ 0.067 & 0.063 $\pm$ 0.127 & 3 & $-$  & $-$ \\
Bootes~III & ab+c & $-$0.048 $\pm$ 0.043 & $-$0.209 $\pm$ 0.187 & 5 & 0.032 $\pm$ 0.049 & 0.060 $\pm$ 0.092 & 5 & $-$  & $-$  \\
\midrule
Sculptor & ab & $-$0.046 $\pm$ 0.056 & $-$0.198 $\pm$ 0.245 & 206 & 0.008 $\pm$ 0.074 & 0.014 $\pm$ 0.139 & 205 & 0.425 $\pm$ 0.361 & 216 \\
Sculptor & c & $-$0.051 $\pm$ 0.062 & $-$0.219 $\pm$ 0.268 & 144 & 0.021 $\pm$ 0.077 & 0.040 $\pm$ 0.145 & 144 & $-$ & $-$ \\
Sculptor & ab+c & $-$0.048 $\pm$ 0.059 & $-$0.207 $\pm$ 0.255 & 350 & 0.013 $\pm$ 0.075 & 0.025 $\pm$ 0.142 & 349 & $-$ & $-$ \\
\midrule
Ursa Minor & ab & $-$0.054 $\pm$ 0.063 & $-$0.235 $\pm$ 0.274 & 45 & 0.017 $\pm$ 0.086 & 0.033 $\pm$ 0.162 & 45 & 0.230 $\pm$ 0.226 & 58 \\
Ursa Minor & c & $-$0.039 $\pm$ 0.064 & $-$0.169 $\pm$ 0.278 & 38 & 0.035 $\pm$ 0.075 & 0.067 $\pm$ 0.142 & 38 & $-$ & $-$ \\
Ursa Minor & ab+c & $-$0.047 $\pm$ 0.064 & $-$0.205 $\pm$ 0.276 & 83 & 0.026 $\pm$ 0.081 & 0.048 $\pm$ 0.153 & 83 & $-$ & $-$ \\
\bottomrule
\end{tabular}
\tablefoot{ Results are given separately for RRab, RRc, and the combined sample (RRab+RRc).  $N$ is the number of stars used for each estimate. 
In systems where only one RR Lyrae star is available, the reported uncertainty corresponds to the individual error. For comparison, we also list the mean $A_{G_{DR3}}$ values derived from the individual absorption estimates published in the {\it Gaia} DR3 {\tt vari\_rrlyrae} table.}
\end{table*}


\begin{table*}[h]
\tiny
\centering
\caption{ Same as in Table~\ref{tab:all_sistemi} but for GCs. Only a portion of the table is shown here for guidance regarding its form and content. 
}
\label{tab:all_gcs}
\begin{tabular}{llcclcclcl}
\toprule
System & Type & E($G-G_{RP}$) & A$_{G_{(G-G_{RP})}}$ & N & E($G_{BP}-G_{RP}$) & A$_{G_{(G_{BP}-G_{RP})}}$ & N & A$_{G_{DR3}}$ & N\\
& & (mag) & (mag) & & 
(mag) & (mag) &  & (mag) & \\
\bottomrule
Arp2 & ab & 0.068 $\pm$ 0.029 & 0.294 $\pm$ 0.128 & 2 & 0.144 $\pm$ 0.068 & 0.272 $\pm$ 0.129 & 2 & 0.345 $\pm$ 0.181 & 3 \\
Arp2 & c & 0.157 $\pm$ 0.028 & 0.679 $\pm$ 0.121 & 1 & 0.206 $\pm$ 0.051 & 0.390 $\pm$ 0.096 & 1 & $-$ & $-$ \\
Arp2 & ab+c & 0.097 $\pm$ 0.055 & 0.423 $\pm$ 0.240 & 3 & 0.165 $\pm$ 0.060 & 0.311 $\pm$ 0.114 & 3 & $-$ & $-$ \\
\midrule
Djorg2 & ab & 0.506 $\pm$ 0.075 & 2.196 $\pm$ 0.326 & 1 & 1.128 $\pm$ 0.105 & 2.130 $\pm$ 0.193 & 1 & 2.479 $\pm$ 0.365 & 1 \\
Djorg2 & c & $-$ & $-$ & $-$ & $-$ & $-$ & $-$ & $-$ & $-$ \\
Djorg2 & ab+c & 0.506 $\pm$ 0.075 & 2.196 $\pm$ 0.326  & 1 & 1.128 $\pm$ 0.105 & 2.130 $\pm$ 0.193 & 1 & $-$ & $-$ \\
\midrule
IC4499 & ab & 0.157 $\pm$ 0.033 & 0.681 $\pm$ 0.143 & 42 & 0.373 $\pm$ 0.070 & 0.704 $\pm$ 0.132 & 42 & 0.600 $\pm$ 0.119 & 44 \\
IC4499 & c & 0.153 $\pm$ 0.033 & 0.663 $\pm$ 0.145 & 17 & 0.328 $\pm$ 0.033 & 0.621 $\pm$ 0.062 & 17 & $-$ & $-$ \\
IC4499 & ab+c & 0.156 $\pm$ 0.033 & 0.676 $\pm$ 0.143 & 59 & 0.360 $\pm$ 0.065 & 0.680 $\pm$ 0.122 & 59 & $-$ & $-$ \\
... &... & ... &... & ... & ... & ...& ... & ... & ...\\
\bottomrule
\end{tabular}
\tablefoot{The full table with all RR Lyrae stars in Galactic GCs is available in electronic form at the CDS.}
\end{table*}
\begin{table}
\centering
\caption{Mean value and standard deviation of the $E(B-V)$ color excess for the classical and ultra–faint dSph galaxies in our sample (RRab+RRc stars combined).}
\label{tab:comparison_sistemi}
\footnotesize 
\setlength{\tabcolsep}{4pt} 
\begin{tabular}{lccc}
\hline
System & $E(B-V)_{(G-G_{RP})}$ & $E(B-V)_{(G_{BP}-G_{RP})}$ & $E(B-V)_{Lit.}$ \\
       & (this work)       & (this work)         & (Literature)    \\
\midrule
Carina$^a$    & $-0.036 \pm 0.298$ & $0.060 \pm 0.147$  & $0.03 \pm 0.02$ \\
Coma$^b$      & $0.006 \pm 0.083$  & $0.016 \pm 0.086$  & $0.045 \pm 0.015$ \\
Draco$^c$     & $-0.046 \pm 0.130$ & $0.051 \pm 0.085$  & $0.027$           \\
Hydrus I$^d$  & $0.165 \pm 0.077$  & $0.114 \pm 0.032$  & $\sim0.097$       \\
Sextans$^e$   & $0.004 \pm 0.198$  & $0.094 \pm 0.155$  & $0.02 \pm 0.02$   \\
Tucana II$^d$ & $-0.131 \pm 0.075$ & $0.025 \pm 0.067$  & $\sim0.019$       \\
Tucana III$^d$& $-0.019 \pm 0.060$ & $0.047 \pm 0.060$  & $\sim0.012$       \\
UMa I$^f$     & $-0.083 \pm 0.088$ & $0.025 \pm 0.069$  & $0.04 \pm 0.02$   \\
UMa II$^g$    & $0.155 \pm 0.078$  & $0.180 \pm 0.066$  & $0.096$           \\
Bootes I$^h$  & $-0.026 \pm 0.087$ & $0.043 \pm 0.047$  & $0.017$           \\
Bootes II$^h$ & $-0.059 \pm 0.086$ & $0.033 \pm 0.054$  & $0.031$           \\
Bootes III$^h$& $-0.083 \pm 0.074$ & $0.024 \pm 0.037$  & $0.021$           \\
Sculptor$^i$  & $-0.082 \pm 0.101$ & $0.010 \pm 0.056$  & $0.018$           \\
Ursa Minor$^j$& $-0.081 \pm 0.110$ & $0.019 \pm 0.061$  & $0.03 \pm 0.01$   \\
\hline
\end{tabular}
\tablefoot{The $A_G$ values were converted into $E(B-V)$ using $A_G / R_G$ with $R_G = 2.516 \pm 0.036$. \\
\textbf{References:} 
($a$) \citet{Mateo-1998,Dallora-2003,Coppola-2015}; 
($b$) \citet{Musella-2009}; 
($c$) \citet{Bonanos-2004}; 
($d$) \citet{Vivas2020b}; 
($e$) \citet{Mateo-1991}; 
($f$) \citet{Garofalo-2013}; 
($g$) \citet{Dallora-2012}; 
($h$) \citet{McConnachie-2012}; 
($i$) \citet{Pietrzynski-2008}; 
($j$) \citet{Garofalo-2025}.
}
\end{table}

\begin{table*}
\centering
\caption{Comparison of the mean $E(B-V)$ values for GCs (RRab+RRc sample).}
\label{tab:comparison_gcs}
\footnotesize
\setlength{\tabcolsep}{2pt} 
\begin{tabular}{l c c c c}
\hline\hline
System & $E(B-V)_{(G-G_{RP})}$ & $E(B-V)_{(G_{BP}-G_{RP})}$ & $E(B-V)_{\rm M14}$ & $E(B-V)_{\rm H96}$ \\
       & (this work) & (this work) & \citealt{McNamara-2014} & \citealt{Harris-1996} \\
\hline
NGC 1466 (LMC)   & $-0.072 \pm 0.121$ & $0.051 \pm 0.062$  & $0.088 \pm 0.002$  & $-$    \\
Reticulum        & $-0.082 \pm 0.103$ & $0.045 \pm 0.061$  & $0.028 \pm 0.004$  & $-$    \\
NGC 1841 (LMC)   & $0.095 \pm 0.091$  & $0.180 \pm 0.028$  & $0.162 \pm 0.012$  & $-$    \\
NGC 2257 (LMC)   & $-0.116 \pm 0.096$ & $0.043 \pm 0.044$  & $0.032 \pm 0.003$  & $-$    \\
NGC 1851         & $0.025 \pm 0.064$  & $0.066 \pm 0.041$  & $0.036 \pm 0.002$  & $0.03$ \\
NGC 3201         & $0.279 \pm 0.067$  & $0.322 \pm 0.061$  & $0.304 \pm 0.002$  & $0.24$ \\
NGC 4590 (M68)   & $0.065 \pm 0.068$  & $0.101 \pm 0.033$  & $0.075 \pm 0.008$  & $0.05$ \\
NGC 5053         & $-0.014 \pm 0.026$ & $0.029 \pm 0.021$  & $0.024 \pm 0.004$  & $0.01$ \\
NGC 5466         & $-0.031 \pm 0.051$ & $0.037 \pm 0.051$  & $-0.005 \pm 0.002$ & $0.00$ \\
NGC 5904 (M5)    & $0.015 \pm 0.099$  & $0.079 \pm 0.047$  & $0.036 \pm 0.002$  & $0.03$ \\
NGC 6266 (M62)   & $0.583 \pm 0.032$  & $0.558 \pm 0.097$  & $0.518 \pm 0.004$  & $0.47$ \\
NGC 6341 (M92)   & $-0.016 \pm 0.083$ & $0.077 \pm 0.056$  & $0.023 \pm 0.002$  & $0.02$ \\
NGC 6388         & $0.503 \pm 0.043$  & $0.411 \pm 0.019$  & 0.4 & $0.37$ \\
NGC 6809 (M55)   & $0.094 \pm 0.102$  & $0.134 \pm 0.074$  & $0.111 \pm 0.004$  & $0.08$ \\
NGC 6864 (M75)   & $0.143 \pm 0.003$  & $0.198 \pm 0.042$  & $0.158 \pm 0.002$  & $0.16$ \\
NGC 7078 (M15)   & $0.157 \pm 0.065$  & $0.131 \pm 0.023$  & $0.110 \pm 0.002$  & $0.1\phantom{0}$  \\
NGC 7089 (M2)    & $0.010 \pm 0.064$  & $0.055 \pm 0.040$  & $0.023 \pm 0.004$  & $0.06$ \\
Rup 106          & $0.238 \pm 0.039$  & $0.214 \pm 0.019$  & $0.142 \pm 0.002$  & $0.2\phantom{0}$  \\
\hline
\end{tabular}
\tablefoot{ Our results are given for the combined RRab+RRc sample.  
$A_G$ values were converted to $E(B-V)$ using $R_G = 2.516 \pm 0.036$.
}
\end{table*}
Based on the PAC(Z) relations derived in the previous section, we estimated the intrinsic $(G-G_{\rm RP})_0$ and $(G_{\rm BP}-G_{\rm RP})_0$ colors for the 
RR Lyrae stars in the {\it Gaia} DR3 {\tt vari\_rrlyrae} catalogue, and derived the corresponding $E(G-G_{\rm RP})$ and $E(G_{\rm BP}-G_{\rm RP})$ color excesses.\\
For the RRab stars, we adopted individual metal abundances from \citet{Muraveva-2025}, which are on the \citet{Crestani-2021} metallicity scale. However, we preliminary converted those metallicities to the  \citet{ZW-1984} scale, using inverse transformations from \citet{Muraveva-2025}, to  render them consistent with the PACZ relations derived in this work (see Section~\ref{sec:data}).\\
The $A_V$ values of our M21 reference samples, were converted into {\it Gaia}-band absorptions ($A_G$, $A_{G_{\rm BP}}$ and $A_{G_{\rm RP}}$) using the total-to-selective extinction ratios from \citet{Huang-2021}. 
For consistency, the color excesses derived for the full RR Lyrae sample were converted into $G$-band absorption values using the same set of ratios.\\
The relations can be written as:
\[
E(G-G_{\rm RP}) = A_G - A_{G_{\rm RP}} = (R_G - R_{G_{\rm RP}})\,E(B-V),
\]
which leads to\[A_G = \frac{R_G}{R_G - R_{G_{\rm RP}}}\,E(G-G_{\rm RP}).\]
Adopting $R_G = 2.516 \pm 0.036$ and $R_{G_{\rm RP}} = 1.936 \pm 0.025$, we obtain
\begin{equation}\label{eq:egrp}
A_G \simeq (4.340 \pm 0.083)\,E(G-G_{\rm RP}),
\end{equation}
Similarly, using the $(G_{\rm BP}-G_{\rm RP})$ color excess:
\[
A_G = \frac{R_G}{R_{G_{\rm BP}}-R_{G_{\rm RP}}}\,E(G_{\rm BP}-G_{\rm RP}),
\]and adopting $R_{G_{\rm BP}} = 3.266 \pm 0.037$, we obtain
\begin{equation}\label{eq:ebprp}
A_G \simeq (1.889 \pm 0.038)\,E(G_{\rm BP}-G_{\rm RP}).
\end{equation}
For comparison, in \citet{Garofalo-et-al-2022} we adopted the coefficient \(\lambda = A_G / E(G_{\rm BP}-G_{\rm RP}) = 1.922\), 
i.e. a difference \(\Delta\lambda = 1.922 - 1.889 = 0.033\) ($\sim 1.7\%$). 
For a representative color excess of \(E(G_{\rm BP}-G_{\rm RP})=0.10\) mag, this corresponds to a change in $A_G$ of only 0.0033 mag, well below the typical uncertainties on individual $A_G$ values in our sample. Therefore, adopting the \citet{Huang-2021} ratios for the $A_V \to A_{\it{Gaia}}$ conversion is both internally consistent with our reference-sample construction and 
equivalent, within the errors, to using the 
coefficient adopted in \citet{Garofalo-et-al-2022}.
To estimate the uncertainties on the derived extinction values, we propagated the measurement errors of all quantities entering the intrinsic color relations.  
For RRc stars,  $(G-G_{\rm RP})_0$ and $(G_{\rm BP}-G_{\rm RP})_0$ were computed from the period $P$ and the peak-to-peak amplitudes in $G$ and $G_{BP}$ bands.  
For RRab stars, the metallicity term [Fe/H] was also included.\\ 
Considering $f(x)$ the intrinsic-color relation, with $x = \{\mathrm{LogP}, \mathrm{Amp}_G \}$ for RRc, and $x = \{\mathrm{LogP}, \mathrm{Amp}_G (or \ \mathrm{Amp}_{G_{BP}}), \mathrm{[Fe/H]}\}$ for RRab.  
The error on $f$, $\sigma_f$, was obtained via standard error propagation:
\begin{equation}
\sigma_f^2 = \sum_i \left( \frac{\partial f}{\partial x_i} \sigma_{x_i} \right)^2,
\end{equation}
where $\sigma_{x_i}$ are the measurement uncertainties on the input parameters.  
The errors on the observed colors, $\sigma_{\rm obs}$, were then combined with $\sigma_f$ to derive the uncertainties on the $E(G-G_{\rm RP})$ and $E(G_{\rm BP}-G_{\rm RP})$ color excesses:
\begin{equation}
\sigma_E^2 = \sigma_{\rm obs}^2 + \sigma_f^2.
\end{equation}
Finally, the uncertainties on $A_G$, were computed by propagating both the errors on the color excesses and the uncertainties on the linear conversion coefficients $k$ derived from the total-to-selective extinction ratios:
\begin{equation}
\sigma_{A_G} = \sqrt{ (k \, \sigma_E)^2 + (E \, \sigma_k)^2 },
\end{equation}
where $\sigma_E$ is the error on the color excess $E(G-G_{\rm RP})$ or $E(G_{\rm BP}-G_{\rm RP})$, and $\sigma_k$ is the uncertainty on the corresponding coefficient $k$ (
$4.340 \pm 0.083$ for ($G-G_{\rm RP}$) and $1.889 \pm 0.038$ for ($G_{\rm BP}-G_{\rm RP}$) see equations~\ref{eq:egrp} and~\ref{eq:ebprp}).  
This procedure provides a self-consistent estimate of the uncertainties on $A_G$ that fully accounts for the measurement errors in periods, amplitudes, metallicity (for RRab stars only), observed magnitudes, and the conversion from color excesses to {\it Gaia}-band extinctions, for both RRc and RRab stars.\\
 To minimize the impact of blending and crowding on the samples used to apply our calibrations, we applied a quality cut to the {\it Gaia} DR3 {\tt vari\_rrlyrae} catalog.  In dense environments like GCs and dwarf galaxies, unresolved background light can suppress pulsation amplitudes. When using amplitude-dependent relations, this suppression artificially reduces the computed intrinsic colors, leading to overestimated color excesses. To mitigate this bias, we calculated the corrected $G_{\rm BP}$ and $G_{\rm RP}$ flux excess factor, $C^*$, as defined by \citet{Riello-2021}\footnote{The corrected flux excess factor is defined as $C^* = C - f(G_{\rm BP} - G_{\rm RP})$, where $C$ is the standard \texttt{phot\_bp\_rp\_excess\_factor} provided in {\it Gaia} DR3, and $f$ is a polynomial function that accounts for the empirical color dependence of the photometric system in normal stars. A threshold of $C^* \le 0.1$ effectively isolates sources with clean, uncontaminated photometry, removing objects affected by local background overestimation or stellar crowding.}, and discarded contaminated sources by applying a threshold of $C^* \le 0.1$ \citep[see also][]{DambisBerdnikov2025}.\\
In the following we have examined the resulting extinction and color-excess distributions across different Galactic environments.

\subsection{All sky}
Figure~\ref{fig:allsky_ag_bprpcu5} shows the all-sky distribution of $A_G$ values for RR Lyrae stars in the {\it Gaia} DR3 {\tt vari\_rrlyrae} catalog, derived from the $(G-G_{\rm RP})$ and $(G_{\rm BP}-G_{\rm RP})$ PAC(Z) relations in Table~\ref{tab:regression_formulas} (top and bottom panels, respectively).  In these maps, the lower limit of the colorbars is set to zero for physical clarity. However, individual stellar estimates providing negative values due to uncertainties in the extinction derivation are fully retained in the dataset to avoid artificial overestimations of the mean values (see Section~\ref{sec:ris_sistemi}). The final analyzed sample includes a total of 142,040 stars for the $(G-G_{\rm RP})$ calibration (90,802 RRab and 51,238 RRc, containing 20,494 RRab and 8,962 RRc with negative absorption), and 141,153 stars for the $(G_{\rm BP}-G_{\rm RP})$ relations (89,915 RRab and 51,238 RRc, with only 2,170 RRab and 2,060 RRc providing $A_G < 0$).
The propagated uncertainties on the color excesses for RRab stars are typically $\sim$0.04--0.07 mag, leading to $A_G$ errors of $\sim$0.13--0.18 mag. For RRc stars, the color-excess uncertainties are $\sim$0.03--0.06 mag, resulting in $A_G$ errors of $\sim$0.12--0.14 mag due to the conversion coefficients.\\
The large–scale structures clearly trace the Galactic disc and bulge, where extinction reaches the highest values, while the halo population appears more uniformly distributed at lower extinction values. The MCs are also clearly visible as overdensities with relatively low and uniform extinction values (see Sect.~\ref{sec:ris_mc}).
 At a qualitative level, the all-sky extinction maps reproduce the expected Galactic extinction pattern, showing an overall consistency between the two sets of relations. The calibrations derived effectively minimise systematic discrepancies, providing excellent agreement between the estimates based on $(G-G_{\rm RP})$ and $(G_{\rm BP}-G_{\rm RP})$ in both classes of pulsation.\\
To quantify the residual difference, we compare the two $A_G$ estimates in Fig.~\ref{fig:gc_confronto_colori}, showing $A_{G_{(G_{\rm BP}-G_{\rm RP})}}$ versus $A_{G_{(G-G_{\rm RP})}}$ for RRab (left panels) and RRc (right panels) stars, respectively, with points color-coded by mean $G$-band magnitude (lower panels) and  number of clean photometric G-band epochs (\texttt{num\_clean\_epochs\_g}; upper panels). 

This quantitative comparison confirms the high stability of the derived calibrations. For the RRc sample, the two methods show a tight convergence, resulting in a negligible median offset of $\sim -0.01$~mag and a remarkably low dispersion ($\sigma = 0.33$~mag). Even at the faint end ($G \approx 18$~mag), the agreement remains solid, with a small median offset of $\sim -0.13$~mag.
The RRab stars exhibit an equally stable and well-behaved trend, displaying a 
uniform median offset of $\sim -0.12$~mag ($\sigma = 0.29$~mag) over the entire magnitude range, which 
shifts to $\sim -0.18$~mag for sources fainter than $G \approx 18$~mag.
Nonetheless, for both classes, the comparison shows a much tighter agreement at the bright end, while the statistical scatter expectedly increases toward fainter magnitudes due to the larger photometric uncertainties.

The overall convergence of the two methods shows the robustness of the empirical relations derived in this work. The $(G_{\rm BP}-G_{\rm RP})$ color remains intrinsically more effective for fine extinction estimations over wide spatial scales primarily because it spans a very wide wavelength range, from approximately 330 to 1050 nm. This baseline is about twice as large as the one covered by $(G-G_{\rm RP})$, providing a higher sensitivity to interstellar dust absorption. However, thanks to the cleaner sample selection achieved through the $C^* \le 0.1$ quality cut applied to the {\tt vari\_rrlyrae} catalog (as detailed in Sect.~\ref{sec:ris}), the $(G-G_{\rm RP})$ relation acts as a perfectly viable and robust alternative, effectively anchoring the $A_G$ estimates even over the shorter wavelength baseline.\\
This level of consistency is preserved even in high-density regions such as the Galactic bulge, confirming the global reliability of both approaches. The remaining minor statistical scatter between individual stellar estimates is likely a reflection of localized data limitations and astrophysical factors. In extremely crowded fields, the $G_{\rm BP}$ and $G_{\rm RP}$ photometry is susceptible to residual crowding and background blending, which can affect the integrated colors of individual sources more than the $G$-band magnitudes. Additionally, secondary contributions to the observed scatter may arise from subtle spatial variations in the interstellar extinction law ($R_V$), which would affect the two color excesses differently along specific lines of sight.
These combined effects, alongside the lower reliability of {\it Gaia} photometry for faint stars ($G \gtrsim 19$), suggest that care should be taken when interpreting absolute absorption values. 
In the following section, we validate our $A_G$ estimates using stellar systems where homogeneous reddening is expected. Despite these local limitations, the overall agreement of large-scale features confirms that our relations provide a consistent and homogeneous extinction dataset for {\bf $\gtrsim 140,000$} RR Lyrae stars.\\
For completeness, Fig.~\ref{fig:allsky_ag_bprpcu7} in the Appendix~\ref{sec:appendix_b} shows the analogous maps obtained with the $(G-G_{\rm RP})$ and $(G_{\rm BP}-G_{\rm RP})$ relations, where $G_{\rm BP}$ and $G_{\rm RP}$ are taken from the \texttt{int\_average\_bp} and \texttt{int\_average\_rp} magnitudes provided in the \texttt{vari\_rrlyrae} table of {\it Gaia} DR3. This figure demonstrates that changing the adopted photometry does not qualitatively affect the resulting extinction maps.\\

\subsubsection{Comparison with \citet{SEF2011} extinction maps}
To assess the consistency of our estimates with established literature, we compare our A$_{G}$ values with the maps provided by \citet{SEF2011} (hereafter SF11).
These maps represent a significant recalibration of the infrared dust maps by \citet{Schlegel-1998}, correcting their previously reported 14--25\% overestimation of extinction \citep{SEF2011}. By providing more accurate reddening values and significantly reduced systematic uncertainties, the SF11 maps have become the standard integrated dust maps reference for Galactic studies.
We converted the SF11 $E(B-V)$ values into A$_{G}$ using the R$_{G}$ coefficient from \citet{Huang-2021} discussed in Sect~\ref{sec:data}.
As shown in Fig.~\ref{fig:com_allsky_ag_sef} in Appendix~\ref{sec:appendix_c}, our results exhibit a high degree of global consistency with the SF11 maps across the filtered common sample of 112,584 for the $(G-G_{\rm RP})_0$ calibration and 136,923 for $(G_{\rm BP}-G_{\rm RP})_0$. 
However, the discrepancies increase notably, exceeding 2~mag ($|\Delta A_G| = |A_{G_\text{this work}} - A_{G_{SF11}}|$), as we move toward the Galactic bulge and the thin disk.
As shown in Fig.~\ref{fig:com_allsky_ag_sef_2mag}, where only sources with $|\Delta A_G|>2$~mag are plotted, thanks to the quality cuts applied to our catalog, the number of stars exceeding this threshold is low: only 3,295 sources (2.93\% of the sample) for the $(G-G_{\rm RP})$ calibration, and an equally negligible 2,868 sources (2.09\%) when using the preferred $(G_{\rm BP}-G_{\rm RP})$ relation.
Near the Galactic disk, our A$_{G}$ values are up to 5~mag higher than those of SF11. Conversely, in the innermost bulge, the SF11 maps provide extinction values that are 10--15~mag higher than our RR Lyrae-based estimates.
These localized discrepancies, particularly in the most obscured regions of the bulge, highlight the inherent challenges of mapping extinction where the dust distribution is most complex.\\ Nevertheless, the fact that less than 3\% of our final sample shows a mismatch larger than 2~mag against SF11 confirms the overall stability of our empirical relations.

\subsubsection{Validation via the $M_G$--[Fe/H] relation}
Before exploring our results in GCs and dSph galaxies, we first tested the impact of our extinction estimates on the scatter of the $M_G$--[Fe/H] relation using a select sample of bright all-sky RR Lyrae stars.
We considered the subset of stars analysed in \citet{Garofalo-et-al-2022}, excluding those used as calibrators in our M21 subsamples to ensure independence from the PAC(Z) calibration (Sect.~\ref{sec:data}).
We selected 80 stars among RRab and RRc stars with positive parallaxes, relative parallax uncertainties smaller than 20\%, and valid absorption values. A correction of $-0.033$~mas was applied to the \textit{Gaia} DR3 parallaxes to account for the global zero-point offset \citep{Garofalo-et-al-2022}. Absolute magnitudes in the $G$ band ($M_{G}$) were computed from parallaxes and observed mean $G$ magnitudes without any extinction corrections.\\
 Figure~\ref{fig:lz} displays the comparison of the $M_{G}$ magnitudes as a function of [Fe/H], showing the impact of different extinction corrections arranged in four vertical panels (from top to bottom): (i) $A_{G}$ converted from $A_{V}$ values of SF11 (red points), 
(ii) DR3 values from the \texttt{vari\_rrlyrae} table \citep{Clementini-23} (purple points), 
(iii) our $(G-G_{\rm RP})_0$ relation (green points), and 
(iv) our $(G_{\rm BP}-G_{\rm RP})_0$ relation (blue points). 
In each panel, the uncorrected $M_{G}$ magnitudes (gray points) are shown as a reference. To assess the reliability of each correction, the resulting $M_{G0}$ values are compared with the empirical $M_{G}$--[Fe/H] relation, eq.~19 from \citet{Garofalo-et-al-2022}, $M_{G} = 1.05 + 0.33 \,[\mathrm{Fe/H}]$.
The residuals with respect to this reference line, along with their corresponding standard deviations ($\sigma$), are displayed in the bottom sub-panel of each main panel.

We find that the residuals are smallest when using our PAC(Z)-derived $E(G_{\rm BP}-G_{\rm RP})$ and $E(G-G_{\rm RP})$ relations, with standard deviations of $\sigma \simeq 0.15$--0.17 mag. Slightly larger residuals are obtained using the DR3 absorptions ($\sigma \simeq 0.18$ mag), while the largest dispersion is found when adopting the SF11 
absorption ($\sigma \simeq 0.20$ mag). This indicates that our relations provide a more internally consistent correction of the $M_G$ for both RRab and RRc stars compared to existing literature values.

\subsection{GCs, classical and ultra-faint dSph galaxies}\label{sec:ris_sistemi}

Beside individual extinction values for over 140,000 RR Lyrae stars, we computed also mean $A_G$ and color excesses values for 
a number of well-studied GCs and dSph galaxies (including classical and ultra-faint systems) where members are expected to share a common extinction. For each system, we report in Table~\ref{tab:all_sistemi} (dSphs) and \ref{tab:all_gcs} (GCs) our   
new $E(G-G_{\rm RP})$, $E(G_{\rm BP}-G_{\rm RP})$, and $A_G$ estimates, comparing them with the {\it Gaia}~DR3 {\tt vari\_rrlyrae} values to test consistency and identify potential systematics.
The tables include separate statistics for RRab, RRc, and combined (RRab+RRc) samples. 
\\
To ensure unbiased results, as said before, we did not impose non-negativity constraints on the individual $A_G$ and color excess estimates. In systems with very low intrinsic extinction, photometric scatter can lead to small negative values. Retaining these negative estimates is necessary to avoid a systematic overestimation of the mean extinction. All individual values have therefore been included in the calculation of the system averages.

\begin{figure*}[ht!]
\includegraphics[width=8.9cm]{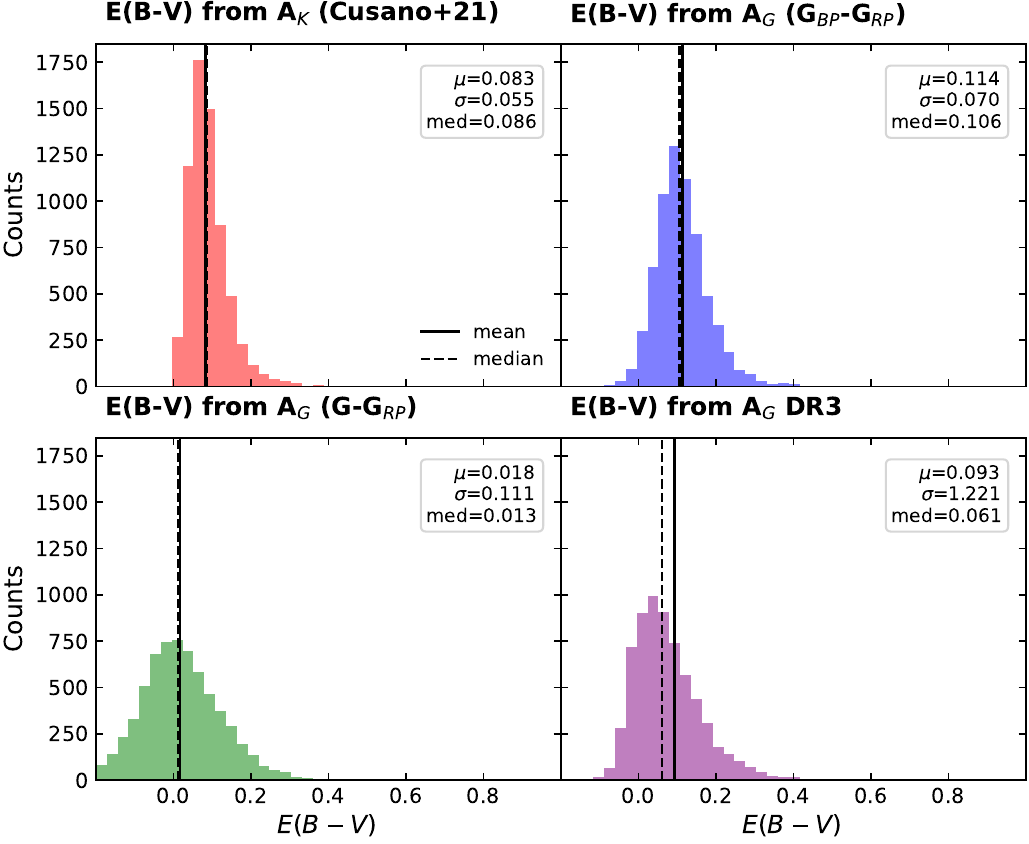}
\includegraphics[width=8.9cm]{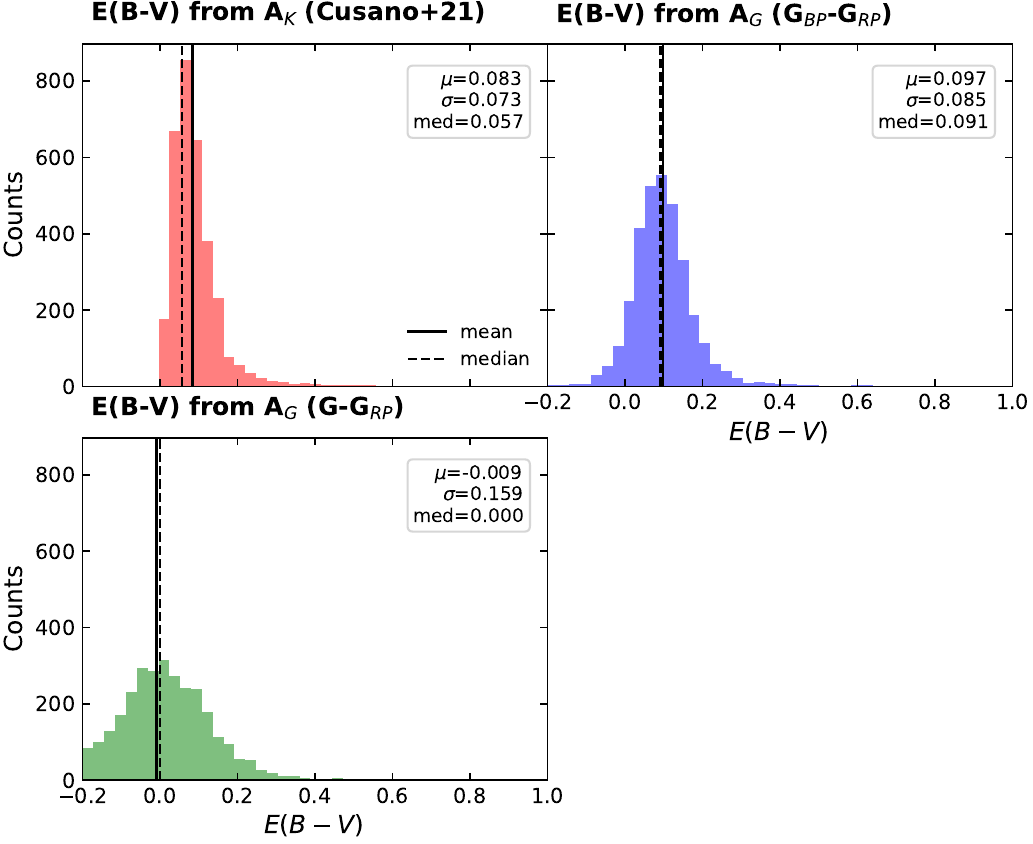}
 \caption{Distributions of $E(B-V)$ for LMC RR Lyrae stars from \citet{Cusano-2021} dataset. 
The four left panels show RRab, and the three right panels show RRc stars.
 Histograms correspond to the distributions derived from the four extinction estimates, including both positive and negative values for the same sample of stars (6605 RRab and 3223 RRc): $A_K$ (red), A$_{G_{(G_{BP}-G_{RP})}}$ (blue; this work), A$_{G_{(G-G_{RP})}}$ (green; this work), and A$_{G_{DR3}}$ (purple).
  }
    \label{fig:ebv_lmc}
\end{figure*}


\subsubsection{Classical and ultra–faint dSph galaxies}
 The comparison of the mean extinction and color excess values obtained for the selected dSph galaxies provides a stringent test of our calibrations in low-reddening regimes. Since these systems are located at high Galactic latitudes in the halo, their expected interstellar dust content is inherently low, with typical literature values of $E(B-V) \lesssim 0.05$ mag (corresponding to $A_G \lesssim 0.13$ mag). Furthermore, because member stars are located at large distances, they lie at the faint end of the {\it Gaia} sensitivity, where photometric scatter increases.\\
The quantitative results summarized in Table~\ref{tab:all_sistemi} show that our empirical relations improve upon the values published in the {\it Gaia} DR3 \texttt{vari\_rrlyrae} catalog ($A_{G_{\rm DR3}}$, hereafter), providing significantly better estimates in several cases. Indeed, the  $A_{G_{\rm DR3}}$ values frequently exhibit greater variations and greater dispersions for these structures, for example, $A_G = 0.913 \pm 1.174$ mag for Carina and $2.744 \pm 2.754$ mag for UMa~I.\\
 When comparing the two empirical calibrations, the $(G_{\rm BP} - G_{\rm RP})$ relation provides a more robust performance that maintains statistical stability near zero extinction. Conversely, the larger intrinsic scatter of the $(G-G_{\rm RP})$ color can artificially bias the distribution towards formally negative absorption values in very low-reddening regimes, reflecting its higher susceptibility to photometric uncertainties rather than a physical trend.
For sparsely populated ultra–faint galaxies, such as Bootes~I, Bootes~III, or Tucana~III, the small number of stars leads to larger uncertainties, making the comparison less constraining (e.g., $A_G = -0.209 \pm 0.187$ mag for Bootes~III). Conversely, in well-populated systems like Draco, Sculptor, and Ursa Minor, the $(G_{\rm BP} - G_{\rm RP})$ relation provides reliable values that closely align with the expected low-reddening ranges, showing significantly smaller standard deviations than $A_{G_{\rm DR3}}$.\\
For RRc stars, the behavior of the two calibrations is remarkably stable thanks to the applied quality cut. Their mean values show a larger scatter only in systems with very few members (e.g., Carina, where $N=3$), reflecting pure small-sample statistics. In well-populated galaxies, the RRc mean from $(G_{\rm BP}-G_{\rm RP})$ is in excellent agreement with the values derived for RRab stars, as seen in Sculptor ($A_G = 0.040 \pm 0.145$ mag for RRc vs. $0.014 \pm 0.139$ mag for RRab).\\
The detailed comparison with independent literature estimates is presented in Table~\ref{tab:comparison_sistemi}, which reports the $E(B-V)$ values derived from our $A_G$ estimates based on both $(G-G_{\rm RP})$ and $(G_{\rm BP}-G_{\rm RP})$ relations. These values were converted using the relation $E(B-V) = A_G / R_G$ with $R_G = 2.516 \pm 0.036$. The table also includes literature $E(B-V)$ values, giving preference to those derived specifically from RR Lyrae stars to ensure a more consistent comparison. 
 Overall, our analysis shows that the $(G_{\rm BP} - G_{\rm RP})$ calibration provides the most reliable and physically consistent results across different reddening regimes. For most systems, its mean $E(B-V)$ values align well with literature estimates. This precision is maintained in well-populated systems like Sculptor ($0.010 \pm 0.056$ mag vs. $0.018$ mag in literature) and Ursa Minor ($0.019 \pm 0.061$ mag vs. $0.03 \pm 0.01$ mag), as well as in ultra–faint structures like Bootes~I ($0.043 \pm 0.047$ mag vs. $0.017$ mag) and Bootes~III ($0.024 \pm 0.037$ mag vs. $0.021$ mag).\\ Conversely, the $(G-G_{\rm RP})$ relation is strictly limited by photometric noise in these faint targets, frequently providing negative values at high Galactic latitudes. Nevertheless, when the intrinsic reddening is slightly higher, as in the case of UMa~II, both relations can provide useful constraints. For example, in UMa~II we find $E(B-V) = 0.180 \pm 0.066$ mag and $0.155 \pm 0.078$ mag for the $(G_{\rm BP}-G_{\rm RP})$ and $(G-G_{\rm RP})$ calibrations, respectively, against a literature value of $0.096$ mag. Despite the large photometric scatter typical of these faint stars, both estimates remain consistent with the literature within their large uncertainties, although the $(G_{\rm BP} - G_{\rm RP})$ relation remains the preferred choice for global applications.

\subsubsection{GCs}\label{ris_gcs}
The results for GCs (Table~\ref{tab:all_gcs}) confirm and extend the trends seen for the dSph galaxies.\\
The $(G_{\rm BP} - G_{\rm RP})$ relations provide reliable and positive estimates across the entire sample.\\
GCs with sufficient numbers of both RRab and RRc stars show that the two classes are in excellent agreement. For instance, in well-populated clusters like NGC~3201 and NGC~5272 (M3), the mean values for RRab and RRc match perfectly. The comparison with $A_{G_{\rm DR3}}$ values shows mixed behavior. In several clusters, particularly those with moderate to high reddening, the mean $A_G$ derived from both of our relations is consistent within uncertainties with $A_{G_{\rm DR3}}$. For bulge or highly obscured clusters (e.g., Djorg~2, NGC~6139, or NGC~6266), both calibrations successfully track the high reddened environment, matching or slightly improving upon the {\it Gaia} DR3 values. In these high-reddening regions, our empirical calibrations provide a very reliable description of extinction.\\
As also shown by the comparison between our $A_G$ and color excesses with {\it Gaia} DR3 estimates, GCs often show higher and more variable extinction than dSphs. This is expected, since many of them are located in the Galactic bulge or disk, where line-of-sight reddening is complex and strongly position-dependent. \\
As done in Table~\ref{tab:comparison_sistemi} for dSphs, Table~\ref{tab:comparison_gcs} summarizes the comparison between our mean color excesses and literature values for GCs, using as literature reference $E(B-V)$ values compiled and reported in table~3 of \citet{McNamara-2014}.\\
Our values are those from the combined RRab+RRc sample, alongside reference estimates from \citet{McNamara-2014} and the \citet{Harris-1996} catalogue. \citet{McNamara-2014} derives $E(B-V)$ for RR Lyrae in GCs by comparing observed mean $(B-V)$ colors with intrinsic colors calculated from empirical relations based on periods and light curve amplitudes of the RRab stars, following the approach of \citet{Piersimoni-2002}. The intrinsic colors account for the pulsation properties of the stars, and small corrections are applied to remove the cluster reddening. \citet{Harris-1996} measurements are mainly inferred from color–magnitude diagrams of the GCs and can be used as an independent reference. 
 The comparison shows that our results from the $(G_{\rm BP} - G_{\rm RP})$ relation agree well with the literature within uncertainties for most GCs. This stability, particularly for the  RRc stars, benefits from the quality cut applied to the sample. In low-reddening systems, this relation remains stable and positive, while the $(G-G_{\rm RP})$ relation tends to measure lower values, although still consistent within the errors, as seen for NGC~1466, Reticulum, and NGC~2257.
In GCs with high extinction (e.g., NGC~6266 and NGC~6388), both relations successfully detect high reddening environment, but the $(G_{\rm BP} - G_{\rm RP})$ relation tends to provide values closer to published estimates. For example, in NGC~6388 we find $E(B-V) = 0.411 \pm 0.019$ mag, which is in excellent agreement with the value of $0.4$ mag from \citet{McNamara-2014}. This supports the conclusion that the $(G_{\rm BP} - G_{\rm RP})$ relations provide more stable results across all extinction environments.

\subsection{LMC and SMC galaxies}\label{sec:ris_mc} 
\begin{figure}
\includegraphics[width=\columnwidth]{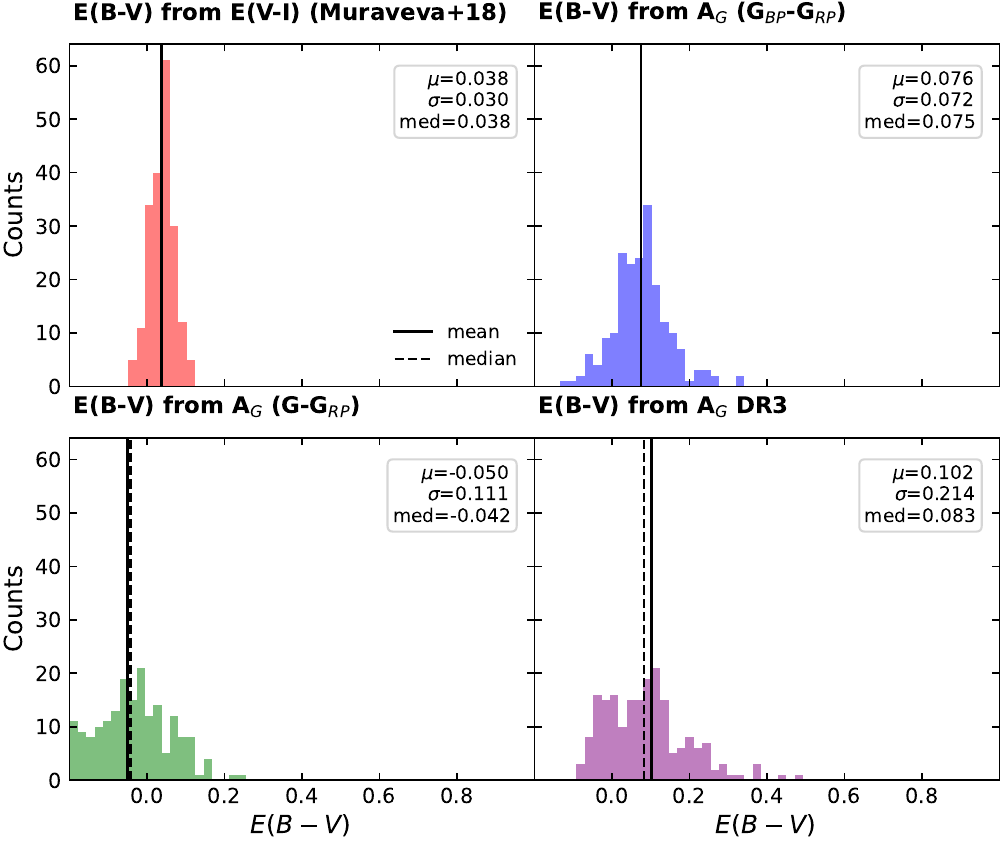}
    \caption{Distributions of $E(B-V)$ for RRab stars in the SMC from the \citet{Muraveva-2018} sample. The histograms shows the full distributions derived from the four extinction estimates, including both positive and negative values for the same sample of stars (198): $E(V-I)$ from \citet{Muraveva-2018} (red), A$_{G_{(G_{BP}-G_{RP})}}$ (blue; this work), A$_{G_{(G-G_{RP})}}$ (green; this work), and A$_{G_{DR3}}$ from {\it Gaia} DR3 \texttt{vari\_rrlyrae} (purple).}
    \label{fig:ebv_smc}
\end{figure}

We next turn to the MCs.\\ 
These systems provide an ideal benchmark for testing extinction estimates, thanks to their well-studied stellar populations, the presence of large number of RR Lyrae stars and the availability of independent reddening determinations from numerous optical and infrared surveys.\\  
In the literature, several studies have addressed the reddening in these systems using different stellar tracers. 
For example, \citet{Haschke-2011} derived reddening maps from OGLE~III data based on RR~Lyrae and red clump stars, finding median values of $E(V-I) = 0.09 \pm 0.07$~mag for the Large Magellanic Cloud (LMC) and $0.04 \pm 0.06$ mag for the Small Magellanic Cloud (SMC), corresponding to $E(B-V) \sim 0.069 \pm 0.054$ mag and $0.031 \pm 0.046$ mag, respectively (using the \citealt{Cardelli-1989} extinction law).\\
More recently, \citet{Skowron-2021} analysed OGLE~IV data and reported mean reddenings of $E(V-I) = 0.100 \pm 0.043$~mag for the LMC and $0.047 \pm 0.025$~mag for the SMC, equivalent to $E(B-V) \approx 0.077 \pm 0.033$~mag and $0.036 \pm 0.019$~mag.  
Independent determinations based on classical Cepheids by \citet{Joshi2019} found $E(V-I) = 0.113 \pm 0.060$~mag for the LMC ($E(B-V) \approx 0.087 \pm 0.046$~mag), while \citet{Deb-2017} estimated $E(B-V) = 0.066 \pm 0.036$~mag for the SMC using red clump stars.  \\
Earlier studies focusing specifically on RR~Lyrae fields, such as \citet{Clementini-et-al-2003}, found $E(B-V) = 0.116 \pm 0.017$~mag and $0.086 \pm 0.017$~mag for two LMC fields (A and B), which were later refined by \citet{McNamara-2014} to $E(B-V) = 0.109 \pm 0.004$~mag and $0.083 \pm 0.003$~mag, respectively.  
Together, these studies outline a consistent picture of relatively low and spatially uniform reddening across both Clouds, typically below $E(B-V) \simeq 0.1$~mag, and provide a robust reference framework for our comparison. \\
We compared our extinction estimates with those derived by \citet{Muraveva-2018} for the SMC and \citet{Cusano-2021} for the LMC. \\
\citet{Muraveva-2018} obtained individual extinction estimates for over 2,900 RRab stars in the SMC, while \citet{Cusano-2021} analysed a much larger sample of about 22,000 RRab and RRc stars in the LMC. Both studies derived optical reddenings, $E(V-I)$, from OGLE~IV $V$ and $I$ photometry using the empirical relation of \citet{Piersimoni-2002}.\\
\citet{Muraveva-2018} provide $E(V-I)$ directly for their RRab stars (mean $\langle E(V-I) \rangle = 0.06 \pm 0.06$ mag, corresponding to $\langle E(B-V) \rangle \approx 0.046 \pm 0.046$ mag), whereas \citet{Cusano-2021} converted the $E(V-I)$ values of RRab stars into $K_s$-band extinctions ($A_{K_s}$) using the standard extinction law \citep{Cardelli-1989}, and built an extinction map to interpolate $A_{K_s}$ for RRc stars and for RR Lyrae lacking $V$ and $I$ magnitudes. 
The mean $K_s$-band extinction of the LMC sample is $\langle A_{K_s} \rangle = 0.03 \pm 0.02$ mag, corresponding to $\langle E(B-V) \rangle \approx 0.086 \pm 0.057$ mag.
To compare these measurements with our $A_G$ values, all extinction estimates were converted into the common scale of color excess $E(B-V)$.
Approximately 90\% of the LMC and SMC RR~Lyrae stars have {\it Gaia}~DR3 counterparts, allowing a direct comparison of the extinction values.  
The distributions of $E(B-V)$ for LMC (RRab and RRc) and SMC (RRab) stars are shown in Figs.~\ref{fig:ebv_lmc} and \ref{fig:ebv_smc}, respectively. These histograms compare 
different estimates: the reference values ($A_K$ from \citealt{Cusano-2021} for the LMC; $E(V-I)$ from \citealt{Muraveva-2018} for the SMC), our results from both $(G_{\rm BP}-G_{\rm RP})$ and $(G-G_{\rm RP})$ colors, and the {\it Gaia}~DR3 \texttt{vari\_rrlyrae} estimates. The complete statistics for these distributions—mean, median, and standard deviation—are detailed in Table~\ref{tab:ebv_comparison_mcs}.
 The statistical properties of the $E(B-V)$ distributions confirm the core trends observed in the previous sections. For the LMC RRab sample (6605 stars), our $(G_{\rm BP} - G_{\rm RP})$ relation provides a mean $E(B-V) = 0.114 \pm 0.070$ mag (median $0.106$ mag), which is slightly higher but fully consistent with the reference values from \citet{Cusano-2021} ($0.083 \pm 0.055$ mag). In contrast, the $(G-G_{\rm RP})$ relation provides a significantly lower mean of $0.018 \pm 0.111$ mag, with a large standard deviation. A similar behavior is found for the LMC RRc sample (3223 stars), where the $(G_{\rm BP} - G_{\rm RP})$ values gives a mean value of $0.097 \pm 0.085$ mag, in agreement the literature estimates much more closely than the values from  $(G-G_{\rm RP})$ ($0.009 \pm 0.159$ mag). Interestingly, the {\it Gaia} DR3 \texttt{vari\_rrlyrae} values show a mean of $0.093$ mag, which matches the literature very well, but they are affected by a severe dispersion ($\sigma = 1.221$ mag).
For the SMC RRab stars, which represent a very low-reddening environment, the reference values from \citet{Muraveva-2018} indicate an average excess of $0.038 \pm 0.030$ mag.  In this system, both of our empirical relations provide results consistent with literature expectations within the uncertainties. The $(G_{\rm BP} - G_{\rm RP})$ relation provides a mean of $0.076 \pm 0.072$ mag, while the $(G-G_{\rm RP})$ relation gives $0.050 \pm 0.111$ mag. On the other hand, the {\it Gaia} DR3 catalogue shows a tendency to overestimate the absorption in this galaxy, providing a higher mean value of $0.102 \pm 0.214$ mag.

Overall, the analysis of the MCs confirms that the $(G_{\rm BP}-G_{\rm RP})$ calibration represents a solid alternative, mitigating the overestimates found in the standard {\it Gaia} DR3 catalog even in low-extinction regimes.

\begin{table}
\centering
\caption{Comparison of $E(B-V)$ estimates for RR~Lyrae in the MCs.}
\label{tab:ebv_comparison_mcs}
\footnotesize
\setlength{\tabcolsep}{3pt}
\begin{tabular}{l l c c c}
\hline
Sample & $E(B-V)$ from & Mean & Median & $\sigma$ \\
       &        & (mag) & (mag) & (mag) \\
\hline
LMC RRab & $A_K$ \citet{Cusano-2021} & 0.083 & 0.086 & 0.055 \\
(6605 stars)      & $A_{G_{(G_{BP}-G_{RP})}}$ (this work)  & 0.114 & 0.106 & 0.070 \\
                  & $A_{G_{(G-G_{RP})}}$ (this work)   & 0.018 & 0.013 & 0.111 \\
                  & $A_{G_{DR3}}$ (\textit{Gaia} DR3)        & 0.093& 0.061 & 1.221 \\
\hline
LMC RRc  & $A_K$ \citet{Cusano-2021} & 0.083 & 0.057 & 0.073 \\
(3223 stars)      & $A_{G_{(G_{BP}-G_{RP})}}$ (this work)  & 0.097 & 0.091 & 0.085 \\
                  & $A_{G_{(G-G_{RP})}}$ (this work)  & 0.009 & 0.000 & 0.159 \\
\hline
SMC RRab & $E(V-I)$ \citet{Muraveva-2018} & 0.038 & 0.038 & 0.030 \\
(198 stars)       & $A_{G_{(G_{BP}-G_{RP})}}$ (this work)   & 0.076 & 0.075 & 0.072 \\
                  & $A_{G_{(G-G_{RP})}}$ (this work)     & 0.050 & 0.042 & 0.111 \\
                  & $A_{G_{DR3}}$ (\textit{Gaia} DR3)           & 0.102 & 0.083 & 0.214 \\
\hline
\end{tabular}
\end{table}

\section{Conclusions}\label{sec:con}
The growing availability of large, homogeneous datasets from {\it Gaia} has enabled new data-driven approaches to study RR~Lyrae stars. In this work, we applied ML techniques to redefine empirical relations and improve extinction estimates, reinforcing the role of RR~Lyrae as reliable reddening indicators. Our calibration is based on a
well-defined Galactic dataset \citep{Muhie-Dambis-2021}, providing a coherent framework for the derivation of both RRab and RRc relations.\\
Specifically, we developed new PACZ relations for RRab stars and, for the first time, analogous PAC 
relations for RRc stars in the {\it Gaia} passbands. The SFS analysis identified period, metallicity, and amplitude as the key predictors to estimate absorption for RRab stars, while RRc stars show a simpler dependence on only two dominant variables (period and amplitude).
The new relations provide intrinsic colors and absolute extinctions with residual scatters of $\sim$0.03--0.04~mag. When applied to the {\it Gaia} DR3 sample after selecting high-quality photometry,  
they provide $A_G$ uncertainties of 0.13--0.18~mag for RRab and 0.12--0.14~mag for RRc stars, demonstrating that the combined effects of photometric uncertainties and regression limits remain well constrained.
Using the PAC(Z)-derived color excesses $E(G-G_{\rm RP})$ and $E(G_{\rm BP}-G_{\rm RP})$, we produced all-sky extinction maps that well trace the Galactic disc, bulge, halo, and the MCs. Comparisons based on a bright field-star sample show that our relations reduce the scatter in the $M_G$–[Fe/H] relation than using the DR3 absorptions, validating their accuracy and suitability for large-scale applications.\\
The effectiveness of these relations was further verified by determining the mean color excesses, $E(G-G_{\rm RP})$ and $E(G_{\rm BP}-G_{\rm RP})$, and the mean absorption $A_G$ for 82 GCs and 14 dSph galaxies. 
 These tests confirm that the empirical relations based on $(G_{\rm BP}-G_{\rm RP})$ provide stable and robust extinction estimates, showing an overall good agreement with literature values across diverse stellar environments, especially for RRab stars. While RRc-based estimates present a slightly higher dispersion in systems sparsely populated by RR Lyrae, they still provide consistent results. Conversely, the $(G-G_{\rm RP})$ relations prove less reliable being more susceptible to larger uncertainties and negative values due to the shorter color baseline.
 On this basis, we recommend adopting the PAC(Z) relations built on $(G_{\rm BP}-G_{\rm RP})$ as more reliable choice for extinction corrections.
Overall, our recalibrated relations for RRab and the first empirical PAC relations for RRc stars in the {\it Gaia} photometric system represent a significant improvement over the {\it Gaia} DR3 implementation and will be instrumental for future studies of stellar populations and Galactic structure.
Looking ahead, the methodology presented here can be further refined. The implementation of non-linear models, such as Random Forest or XGBoost, should be explored to capture potentially more complex relationships between pulsation parameters and intrinsic colors. Such approaches will become increasingly effective as data quality improves with upcoming releases. Specifically, {\it Gaia} DR4, based on 66 months of observations, will offer minimized systematics and enhanced light-curve sampling, providing a firmer basis for investigating non-linear dependencies while ensuring physical interpretability. With these advancements, the relations presented here (and their potential future refinements) are expected to firmly establish RR Lyrae stars as high-precision, independent tracers of interstellar reddening across diverse Galactic and extragalactic environments.

\section*{Data availability}
Tables~\ref{tab:all_gcs} and ~\ref{tab:rrlab_params_full} are only available in electronic form at the CDS via anonymous ftp to cdsarc.u-strasbg.fr (130.79.128.5) or via \url{http://cdsweb.u-strasbg.fr/cgi-bin/qcat?J/A+A/}.

\begin{acknowledgements}
The authors thank the anonymous referee for the constructive comments and suggestions, which helped improve the clarity and robustness of the manuscript.
This work made use of data from the European Space Agency
(ESA) mission Gaia (\url{https://www.cosmos.esa.int/gaia}), processed
by the Gaia Data Processing and Analysis Consortium (DPAC; \url{https://www.cosmos.esa.int/web/gaia/dpac/consortium}). Funding for
the DPAC has been provided by national institutions, in particular,
the institutions participating in the Gaia Multilateral Agreement.
 Support to this study has been provided by INAF Mini-Grant (PI: Tatiana Muraveva), by the Agenzia Spaziale Italiana (ASI) through contract and ASI 2018-24-HH.0 and its Addendum 2018-24-HH.1-2022 and contract ASI 2025-10-H.00, and by Premiale 2015, MIning The Cosmos - Big Data and Innovative Italian Technology for Frontiers Astrophysics and Cosmology (MITiC; P.I.B.Garilli). This research was also supported by the International Space Science Institute (ISSI)
in Bern, through ISSI International Team project 490, ‘SH0T: The
Stellar Path to the H0 Tension in the Gaia, TESS, LSST, and JWST
Era’ (PI: G. Clementini).
 This research also made use of TOPCAT \citep{Taylor-2005} an interactive graphical viewer and editor for tabular data; 
the SIMBAD database, operated at CDS, Strasbourg, France. The original description of the SIMBAD service was published in in \citet{Wenger-2000}; and VizieR catalogue access tool, CDS, Strasbourg, France (DOI:10.26093/cds/vizier). The original description of the VizieR service was published in \citet{Ochsenbein-2000}.
\end{acknowledgements}

%
   \bibliographystyle{aa} 
   \bibliography{pgrex} 
%


\onecolumn
\begin{appendix}

\section{Predicted vs. true intrinsic colors from SFS-based linear models}\label{sec:appendix_a}
This appendix presents diagnostic plots of the SFS-based linear regression analysis for RRab and RRc stars in the M21 subsample.
For each subtype, we show (i) the evolution of the Mean Squared Error (MSE) as a function of the number of selected features, and (ii) the comparison between the predicted and true intrinsic colors. These results provide a visual assessment of the model performance and its stability against overfitting.
The complete lists of parameters used in this work are provided in Tables ~\ref{tab:rrlab_params_full} and~\ref{tab:rrlc_params_full}, for 373 RRab and 38 RRc stars, respectively. 
Column content and origin of each parameter are detailed in the tables caption.

\subsection{RRab stars}\label{sec:appendix_a_rrab}

\begin{figure*}[!ht] 
\sidecaption
  \includegraphics[width=6cm]{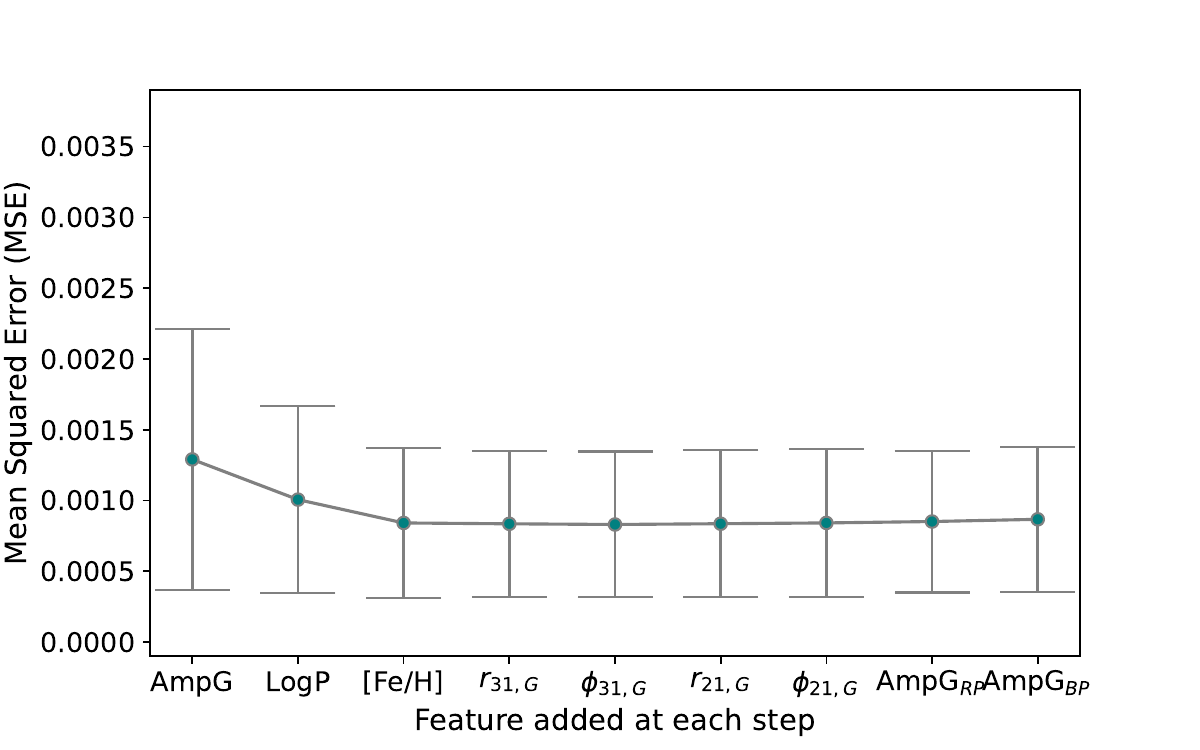}
  \includegraphics[width=6cm]{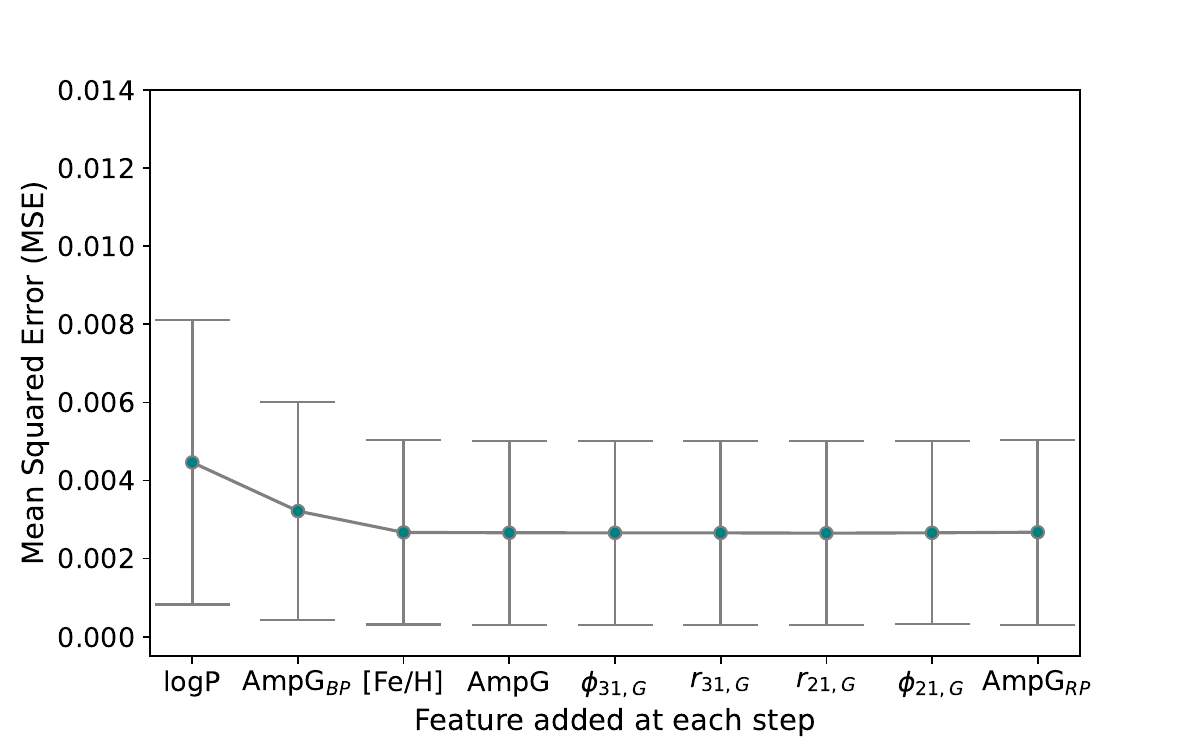} 
  \caption{Evolution of MSE as successive features are added to the linear model for predicting the intrinsic colors $(G-G_{\mathrm{RP}})_0$, left panel, and $(G_{\mathrm{BP}}-G_{\mathrm{RP}})_0$, right panel, for RRab stars in our M21 reference sample. The horizontal axis shows the specific feature added at each step of the SFS procedure.}
  \label{fig:elbow_ab}
\end{figure*}

\begin{figure*}[!ht] 
\sidecaption
  \includegraphics[width=6cm]{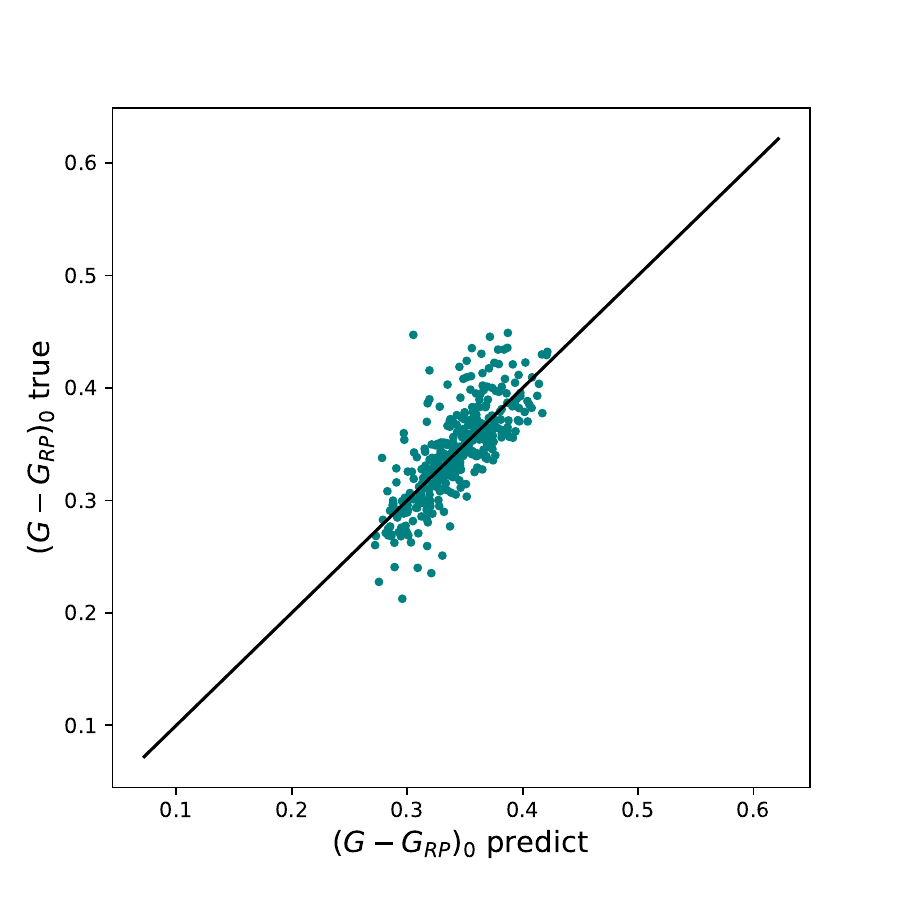}
  \includegraphics[width=6cm]{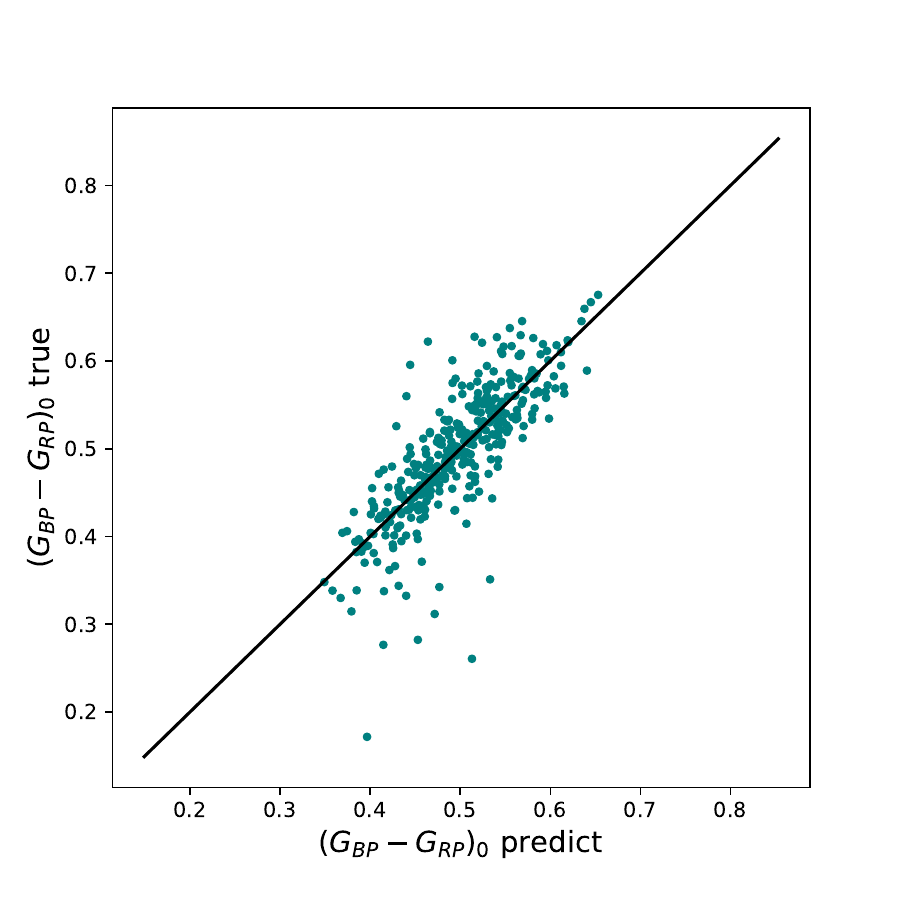}    
  \caption{Comparison between predicted and true intrinsic colors for the RRab stars in the M21 sample. The left panel shows the results for $(G-G_{\mathrm{RP}})_0$, while the right panel displays the comparison for $(G_{\mathrm{BP}}-G_{\mathrm{RP}})_0$. In both panels, the solid black line represents the 1:1 relation. The predictions are based on a linear regression model using features selected via SFS: $\log P$, $\mathrm{[Fe/H]}$ and $\mathrm{Amp}G$ (left panel) and $\log P$, $\mathrm{[Fe/H]}$ and $\mathrm{Amp}G_{\mathrm{BP}}$ (right panel). The root mean square (RMS) of the residuals, which quantifies the model scatter, is 0.027 and 0.044 for $(G-G_{\mathrm{RP}})_0$ and $(G_{\mathrm{BP}}-G_{\mathrm{RP}})_0$, respectively. The relations and the rms values are reported in the upper part of Table~\ref{tab:regression_formulas}.}
  \label{fig:lr_m21_ab}
\end{figure*}

Figures~\ref{fig:elbow_ab} and~\ref{fig:lr_m21_ab} illustrate the results for 
the M21 RRab subsample.
Figure~\ref{fig:elbow_ab} shows the evolution of the MSE with the number of included features for both $(G-G_{\rm RP})_0$ (left panel) and $(G_{\rm BP}-G_{\rm RP})_0$ (right panel). The MSE decreases as features are added, reaching a plateau after three variables, which indicates the optimal feature set.
Figure~\ref{fig:lr_m21_ab} compares the predicted and true intrinsic colors for the same sample. Each panel includes the identity line for reference, showing the overall consistency between model predictions (predict) and observed values (true).
Table~\ref{tab:mse_ab} summarises the performance of the linear regression models used to predict $(G-G_{\rm RP})_0$ and $(G_{\rm BP}-G_{\rm RP})_0$ for RRab stars. The table reports the MSE and the corresponding standard deviation for each set of sequentially added features, highlighting the point at which additional complexity does not significantly improve predictive power.

\begin{table*}[!ht]
\tiny
\centering
\caption{List of parameters used in this work for the 373 RRab stars in our M21 reference sample.}
\label{tab:rrlab_params_full}
\begin{adjustbox}{width=\textwidth, totalheight=\textheight-2cm, keepaspectratio, center} 
\begin{tabular}{lcccccccccccccccc}
\hline
source\_id & RA & Dec & [Fe/H] & $P$ &
$\langle G \rangle$ &
$\langle G_{\rm BP} \rangle$ &
$\langle G_{\rm RP} \rangle$ &
$\mathrm{Amp}G$ &
$\mathrm{Amp{G_{BP}}}$ &
$\mathrm{Amp{G_{RP}}}$ &
$r_{21,G}$ &
$r_{31,G}$ &
$\phi_{21,G}$ &
$\phi_{31,G}$ &
$(G-RP)_0$ &
$(BP-RP)_0$ \\
\hline
5717654459518203392 & 117.583 & $-$18.306 & $-$0.406 & 0.402 & 14.158 & 14.441 & 13.696 & 1.238 & 1.529 & 0.963 & 0.570 & 0.396 & 3.923 & 1.937 & 0.213 & 0.172 \\
5338774988990808064 & 164.742 & $-$59.083 & $-$1.932 & 0.601 & 12.492 & 12.696 & 12.045 & 0.965 & 0.935 & 0.701 & 0.472 & 0.320 & 3.916 & 1.882 & 0.277 & 0.261 \\
3214795601016784256 &  76.977 &  $-$2.147 & $-$1.890 & 0.505 & 12.887 & 12.937 & 12.531 & 1.154 & 1.375 & 0.864 & 0.462 & 0.374 & 3.845 & 1.698 & 0.299 & 0.277 \\
3035123650016214144 & 113.330 & $-$10.489 & $-$1.142 & 0.469 & 13.995 & 14.373 & 13.380 & 0.618 & 1.070 & 0.416 & 0.474 & 0.333 & 3.788 & 1.519 & 0.305 & 0.282 \\
3075045401087186048 & 131.141 &  $-$0.218 & $-$1.552 & 0.537 & 13.303 & 13.491 & 12.953 & 0.947 & 1.173 & 0.712 & 0.476 & 0.359 & 3.906 & 1.866 & 0.251 & 0.312 \\
1818159235220031744 & 310.083 &  22.223 & $-$0.850 & 0.405 & 12.786 & 12.845 & 12.384 & 1.317 & 1.497 & 0.918 & 0.537 & 0.406 & 3.902 & 1.908 & 0.338 & 0.315 \\
4553443379973351296 & 260.661 &  17.885 & $-$1.480 & 0.436 & 12.808 & 12.975 & 12.531 & 1.261 & 1.564 & 0.834 & 0.454 & 0.331 & 3.831 & 1.677 & 0.227 & 0.330 \\
6648425141590751360 & 277.244 & $-$56.233 & $-$1.367 & 0.468 & 13.970 & 14.111 & 13.581 & 1.159 & 1.082 & 0.830 & 0.484 & 0.364 & 3.875 & 1.814 & 0.302 & 0.332 \\
2910611895263048064 &  89.211 & $-$27.667 & $-$1.532 & 0.469 & 12.406 & 12.495 & 12.092 & 1.191 & 1.275 & 0.872 & 0.473 & 0.380 & 3.765 & 1.598 & 0.286 & 0.338 \\
4224859720193721856 & 309.563 &  $-$2.890 & $-$0.580 & 0.362 & 11.787 & 11.933 & 11.454 & 1.105 & 1.431 & 0.867 & 0.549 & 0.333 & 3.891 & 1.912 & 0.271 & 0.338 \\
\dots & \dots & \dots & \dots & \dots & \dots & \dots & \dots & \dots & \dots & \dots & \dots & \dots & \dots & \dots & \dots & \dots \\
\hline
\end{tabular}
\end{adjustbox}
\tablefoot{Columns description:(1) {\it Gaia} DR3 source\_id; (2) and (3) Coordinates; (4) metallicity from M21 on the \citet{ZW-1984} scale; (5) Period; (6)–(8) \texttt{int\_average} $G$, $G_{BP}$ and $G_{RP}$ magnitudes ; (9)–(11) amplitude in $G$, $G_{BP}$ and $G_{RP}$ bands; (12)-(15) Fourier parameters; (16) and (17) $(G-RP)_0$ and $(BP-RP)_0$ values compiled as described in Section~\ref{sec:data}. Columns (1)–(3) are from {\it Gaia} DR3 \texttt{gaia\_source} table \citep{Gaiadr3-2023}. Columns (5)–(15) are from Gaia DR3 \texttt{vari\_rrlyrae} table \citep{Clementini-23}. The full table is available in electronic form at the CDS.
A portion is shown here for guidance regarding its form and content.}
\end{table*}

\begin{table}[ht!]
\tiny
\centering
\caption{ Performance of the linear model predicting $(G-G_{\rm RP})_0$ and $(G_{BP}-G_{\rm RP})_0$ intrinsic colors for RRab stars.
MSE with standard deviation for each set of sequentially added features.}
 \label{tab:mse_ab}
\begin{tabular}{lcc}
\hline
Features & MSE & $\sigma_{\rm MSE}$ \\
\hline
 $(G-G_{\rm RP})_0$&  & \\
 $\mathrm{AmpG}$ & 0.001291 & 0.000922 \\
logP,  $\mathrm{AmpG}$ & 0.001006 & 0.000659 \\
logP,  $\mathrm{AmpG}$, [Fe/H] & 0.000842 & 0.000529 \\
logP, $r_{31, G}$,  $\mathrm{AmpG}$, [Fe/H] & 0.000836 & 0.000516 \\
logP, $\phi_{31, G}$, $r_{31, G}$,  $\mathrm{AmpG}$, [Fe/H] & 0.000831 & 0.000515 \\
logP, $\phi_{31, G}$, $r_{31, G}$, $r_{21, G}$, $\mathrm{AmpG}$, [Fe/H] & 0.000836 & 0.000518 \\
logP, $\phi_{21, G}$, $\phi_{31, G}$, $r_{31, G}$, $r_{21, G}$, $\mathrm{AmpG}$, [Fe/H] & 0.000842 & 0.000520 \\
logP, $\phi_{21, G}$, $\phi_{31, G}$, $r_{31, G}$, $r_{21, G}$, $\mathrm{AmpG}$, $\mathrm{AmpG_{RP}}$, [Fe/H] & 0.000852 & 0.000501 \\
logP, $\phi_{21, G}$,$\phi_{31, G}$, $r_{31, G}$, $r_{21, G}$, $\mathrm{AmpG}$, $\mathrm{AmpG_{BP}}$, $\mathrm{AmpG_{RP}}$, [Fe/H] & 0.000867 & 0.000513 \\
\hline
 $(G_{BP}-G_{\rm RP})_0$&  & \\
logP & 0.004464 & 0.003638 \\
logP, $\mathrm{AmpG_{BP}}$ & 0.003217 & 0.002785 \\
logP, $\mathrm{AmpG_{BP}}$, [Fe/H] & 0.002671 & 0.002358 \\
logP, $\mathrm{AmpG_{BP}}$,[Fe/H], $\mathrm{AmpG}$, & 0.002662 & 0.002353 \\
logP, $\phi_{31, G}$, $\mathrm{AmpG}$,$\mathrm{AmpG_{BP}}$, [Fe/H]& 0.002660 & 0.002348 \\
logP, $\phi_{31, G}$, $r_{31, G}$ , $\mathrm{AmpG}$, $\mathrm{AmpG_{BP}}$, [Fe/H] & 0.002659 & 0.002354 \\
logP, $\phi_{31, G}$, $r_{31, G}$ , $r_{21, G}$ , $\mathrm{AmpG}$, $\mathrm{AmpG_{BP}}$, [Fe/H] & 0.002652 & 0.002351 \\
logP, $\phi_{21, G}$, $\phi_{31, G}$, $r_{31, G}$ , $r_{21, G}$ , $\mathrm{AmpG}$, $\mathrm{AmpG_{BP}}$, [Fe/H] & 0.002660 & 0.002345 \\
logP, $\phi_{21, G}$, $\phi_{31, G}$, $r_{31, G}$ , $r_{21, G}$ , $\mathrm{AmpG}$, $\mathrm{AmpG_{BP}}$, $\mathrm{AmpG_{RP}}$, [Fe/H] & 0.002677 & 0.002373 \\
\hline
\end{tabular}
\end{table}

\subsection{RRc stars}\label{sec:appendix_a_rrc}
\begin{figure*}[!ht]
\sidecaption
  \includegraphics[width=6cm]{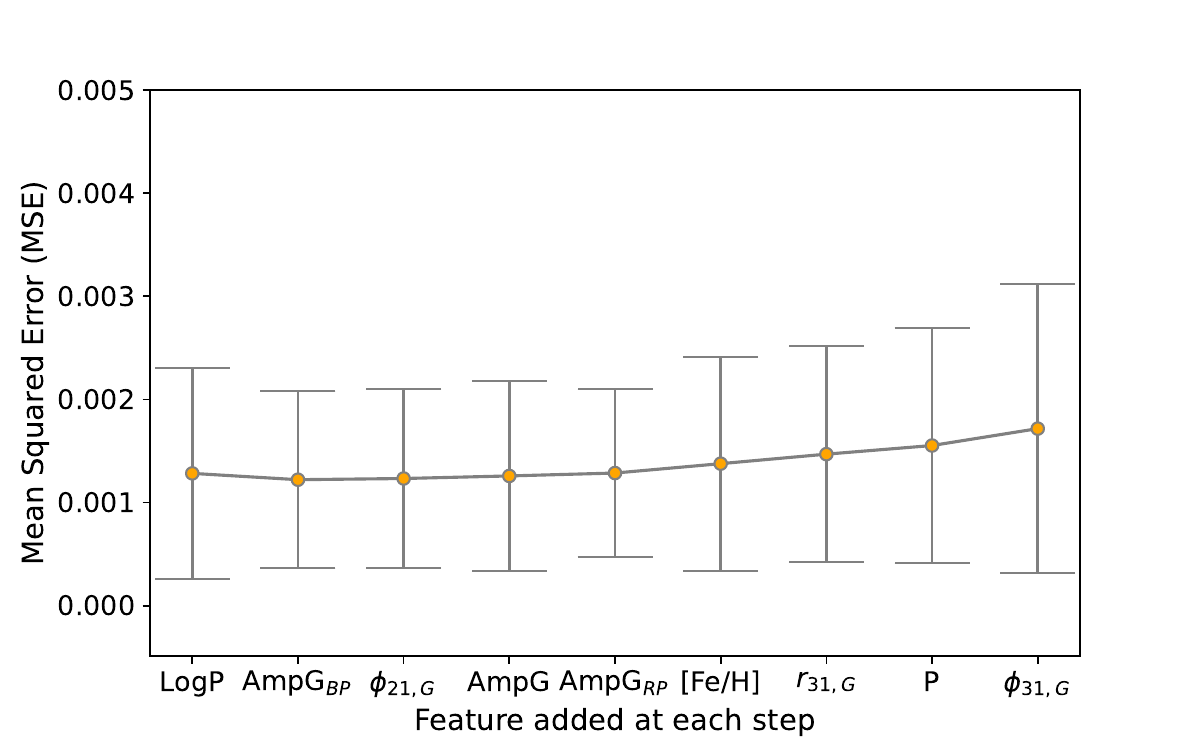}
  \includegraphics[width=6cm]{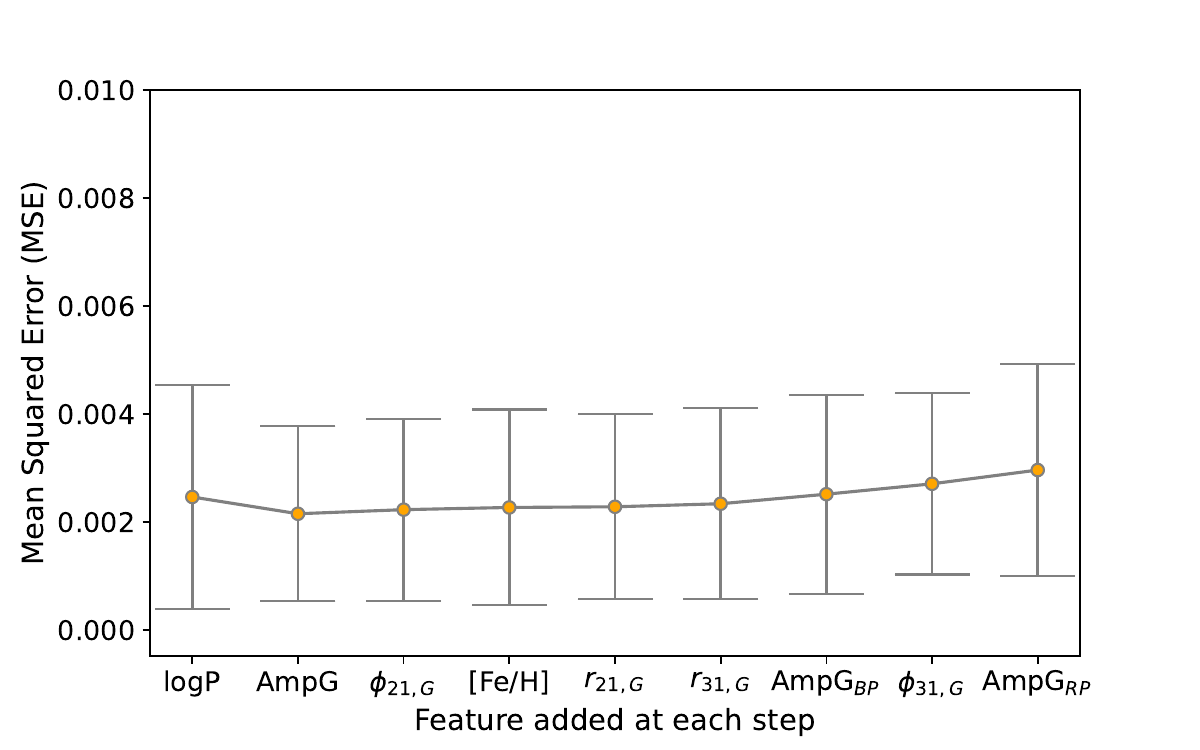}
  \caption{Same as in Figure~\ref{fig:elbow_ab} but for RRc stars. In this case, the MSE does not exhibit a plateau, suggesting that only a limited number of features, two ($\log P$ and $\mathrm{Amp}G$), contribute meaningfully to the prediction of intrinsic color.}
  \label{fig:elbow_c}
\end{figure*}

\begin{figure*}[!ht]
\sidecaption
  \includegraphics[width=6cm]{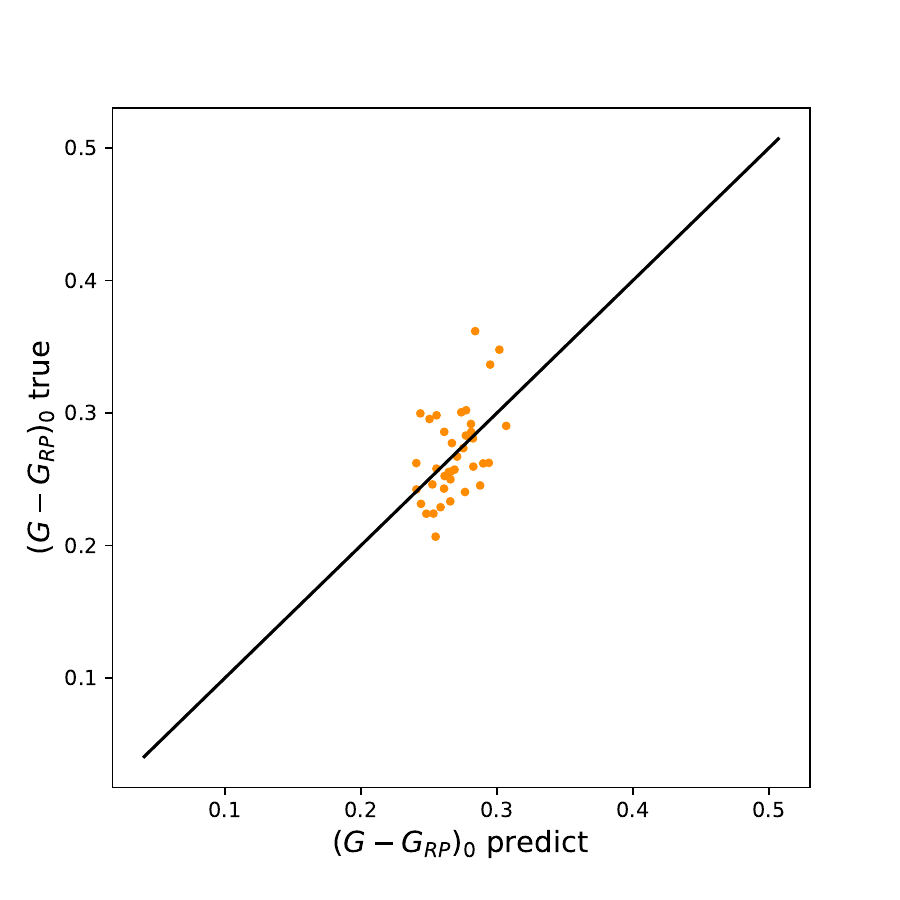}
  \includegraphics[width=6cm]{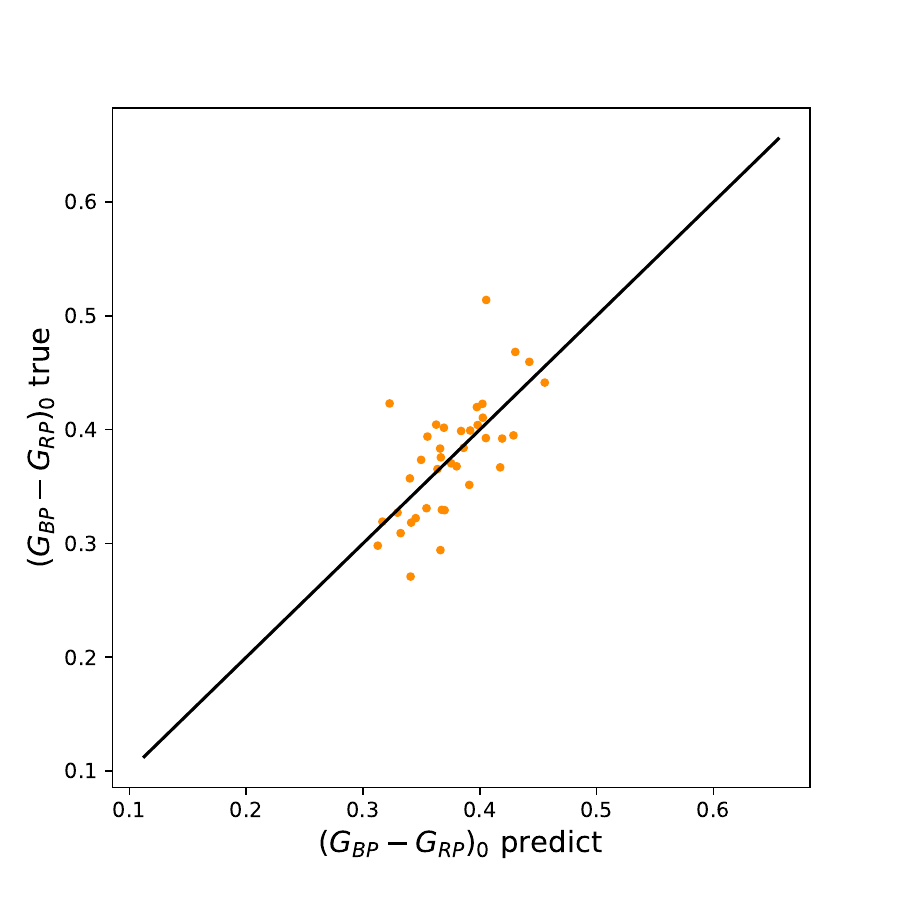}    
  \caption{Same as in Figure~\ref{fig:lr_m21_ab} but for RRc stars. {\it Left panel:} For intrinsic colors $(G - G_{\mathrm{RP}})_0$, predictions are based on a linear regression model using features selected via SFS: $\log P$ and $\mathrm{Amp}G$. The RMS of the residuals is 0.029 (0.029 with $\mathrm{Amp}G_{\mathrm{BP}}$). {\it Right panel:} For $(G_{\mathrm{BP}} - G_{\mathrm{RP}})_0$, the predictions are based on a linear regression model using features selected via SFS: $\log P$ and $\mathrm{Amp}G$ with RMS = 0.037. The relations and the rms values are reported in the lower part of Table~\ref{tab:regression_formulas}.}
  \label{fig:lr_m21_c}
\end{figure*}

Figures~\ref{fig:elbow_c} and~\ref{fig:lr_m21_c} present the corresponding results for RRc stars.
As shown in Figure~\ref{fig:elbow_c}, and quantitatively summarised in Table~\ref{tab:mse_c}, 
the MSE reaches its minimum when only two features are used.
Including additional parameters leads to an increase in MSE, indicating that further variables do not improve 
the predictive performance and may instead introduce noise.
Figure~\ref{fig:lr_m21_c} compares the predicted and true intrinsic colors for the RRc subsample.
The identity line is included for reference, showing good overall agreement within the expected scatter and 
confirming the adequacy of the two–feature regression model. 


\begin{table}[ht!]
\centering
\tiny
\caption{Performance of the linear model for RRc stars predicting $(G-G_{\rm RP})_0$ and $(G_{BP}-G_{\rm RP})_0$.
MSE with standard deviation for each set of sequentially added features.}
\begin{tabular}{lcc}
\hline
Features & MSE & $\sigma_{\rm MSE}$ \\
\hline
 $(G-G_{\rm RP})_0$&  & \\
logP & 0.001283 & 0.001025 \\
logP, $\mathrm{AmpG_{BP}}$ & 0.001222 & 0.000859 \\
logP, $\phi_{21, G}$, $\mathrm{AmpG_{BP}}$ & 0.001233 & 0.000867 \\
logP, $\phi_{21, G}$, $\mathrm{AmpG_{BP}}$, $\mathrm{AmpG_{RP}}$& 0.001258 & 0.000924 \\
logP, $\phi_{21, G}$, $\mathrm{AmpG}$, $\mathrm{AmpG_{BP}}$, $\mathrm{AmpG_{RP}}$ & 0.001286 & 0.000812 \\
logP, $\phi_{21, G}$, $\mathrm{AmpG}$, $\mathrm{AmpG_{BP}}$, $\mathrm{AmpG_{RP}}$, [Fe/H] & 0.001377 & 0.001038 \\
logP, $\phi_{21, G}$, $\mathrm{AmpG}$, $\mathrm{AmpG_{BP}}$, $\mathrm{AmpG_{RP}}$,$r_{31, G}$, [Fe/H] & 0.001469 & 0.001047 \\
logP, P, $\phi_{21, G}$, $\mathrm{AmpG}$, $\mathrm{AmpG_{BP}}$, $\mathrm{AmpG_{RP}}$,$r_{31, G}$, [Fe/H] & 0.001553 & 0.001140 \\
logP, P, $\phi_{21, G}$, $\mathrm{AmpG}$, $\mathrm{AmpG_{BP}}$, $\mathrm{AmpG_{RP}}$, $\phi_{31, G}$,$r_{31, G}$, [Fe/H] & 0.001717 & 0.001404 \\
\hline
 $(G_{BP}-G_{\rm RP})_0$&  & \\
logP & 0.002461 & 0.002073 \\
logP, $\mathrm{AmpG}$ & 0.002150 & 0.001623 \\
logP, $\phi_{21, G}$, $\mathrm{AmpG}$ & 0.002226 & 0.001686 \\
logP, $\phi_{21, G}$, $\mathrm{AmpG}$, [Fe/H] & 0.002266 & 0.001815 \\
logP, $\phi_{21, G}$, $r_{21, G}$, $\mathrm{AmpG}$, [Fe/H] & 0.002279 & 0.001712 \\
logP, $\phi_{21, G}$, $r_{21, G}$, $\mathrm{AmpG}$, $r_{31, G}$, [Fe/H] & 0.002335 & 0.001772 \\
logP, $\phi_{21, G}$, $r_{21, G}$, $\mathrm{AmpG}$, $\mathrm{AmpG_{BP}}$, $r_{31, G}$, [Fe/H] & 0.002512 & 0.001842 \\
logP, $\phi_{21, G}$, $r_{21, G}$, $\mathrm{AmpG}$, $\mathrm{AmpG_{BP}}$, $\mathrm{AmpG_{RP}}$, $r_{31, G}$, [Fe/H] & 0.002704 & 0.001680 \\
logP, $\phi_{21, G}$, $r_{21, G}$, $\mathrm{AmpG}$, $\mathrm{AmpG_{BP}}$, $\mathrm{AmpG_{RP}}$, $\phi_{31, G}$, $r_{31, G}$, [Fe/H] & 0.002959 & 0.001958 \\
\hline
\end{tabular}
 \label{tab:mse_c}
\end{table}

\begin{table*}[!ht]
\tiny
\centering
\caption{List of parameters used in this work for the 38 RRc stars in our M21 reference sample. Column content and origin for each parameter are as in Table~\ref{tab:rrlab_params_full} for the RRab sample.}
\label{tab:rrlc_params_full}
\begin{adjustbox}{width=\textwidth, totalheight=\textheight-2cm, keepaspectratio, center} 
\begin{tabular}{lcccccccccccccccc}
\hline
source\_id & RA & Dec & [Fe/H] & $P$ &
$\langle G \rangle$ &
$\langle G_{\rm BP} \rangle$ &
$\langle G_{\rm RP} \rangle$ &
$\mathrm{Amp}G$ &
$\mathrm{Amp{G_{BP}}}$ &
$\mathrm{Amp{G_{RP}}}$ &
$r_{21,G}$ &
$r_{31,G}$ &
$\phi_{21,G}$ &
$\phi_{31,G}$ &
$(G-RP)_0$ &
$(BP-RP)_0$ \\
\hline
1134921885080388992 & 119.745 & 72.788 & $-$1.420 & 0.267 & 11.471 & 11.560 & 11.245 & 0.313 & 0.409 & 0.227 & 0.115 & 0.034 & 4.683 & 3.126 & 0.207 & 0.271\\
5382867123212474624 & 172.769 & $-$41.118 & $-$1.990 & 0.353 & 12.394 & 12.601 & 12.031 & 0.559 & 0.674 & 0.425 & 0.293 & 0.100 & 4.337 & 2.043 & 0.243 & 0.294\\
6246499215814759936 & 238.965 & $-$21.809 & $-$0.762 & 0.254 & 11.431 & 11.696 & 10.991 & 0.349 & 0.424 & 0.258 & 0.099 & 0.068 & 4.746 & 3.745 & 0.262 & 0.298\\
4643476198121275648 & 38.328 & $-$73.612 & $-$1.282 & 0.287 & 11.976 & 12.092 & 11.728 & 0.428 & 0.594 & 0.342 & 0.189 & 0.052 & 4.833 & 2.713 & 0.224 & 0.309\\
3962473193054146944 & 190.013 & 27.499 & $-$1.360 & 0.293 & 11.680 & 11.763 & 11.426 & 0.421 & 0.528 & 0.345 & 0.172 & 0.168 & 5.208 & 3.271 & 0.246 & 0.318\\
6884361748289023488 & 319.885 & $-$15.117 & $-$1.220 & 0.273 & 11.217 & 11.374 & 10.913 & 0.423 & 0.498 & 0.322 & 0.172 & 0.086 & 4.708 & 2.825 & 0.242 & 0.319\\
1565435491138161664 & 201.556 & 56.257 & $-$1.780 & 0.307 & 10.789 & 10.895 & 10.560 & 0.464 & 0.589 & 0.338 & 0.215 & 0.059 & 4.664 & 2.936 & 0.224 & 0.322\\
1492230556717187456 & 214.152 & 42.360 & $-$2.240 & 0.313 & 10.914 & 11.017 & 10.677 & 0.538 & 0.628 & 0.399 & 0.317 & 0.084 & 4.323 & 2.369 & 0.231 & 0.327\\
6183829286409316480 & 193.016 & $-$31.048 & $-$1.585 & 0.323 & 13.214 & 13.366 & 12.938 & 0.440 & 0.545 & 0.339 & 0.197 & 0.080 & 4.838 & 2.962 & 0.233 & 0.329\\
5242353351117203200 & 150.349 & $-$71.284 & $-$1.233 & 0.304 & 12.998 & 13.204 & 12.641 & 0.376 & 0.452 & 0.280 & 0.117 & 0.094 & 4.672 & 3.516 & 0.256 & 0.329\\
2852346261548304896 & 355.568 & 24.916 & $-$1.770 & 0.306 & 11.808 & 11.936 & 11.559 & 0.431 & 0.499 & 0.311 & 0.226 & 0.046 & 4.533 & 2.447 & 0.229 & 0.331\\
4918030715504071296 & 1.026 & $-$59.485 & $-$1.250 & 0.333 & 11.062 & 11.181 & 10.816 & 0.406 & 0.481 & 0.299 & 0.095 & 0.043 & 4.745 & 3.637 & 0.240 & 0.351\\
2211629018927324288 & 339.805 & 64.858 & $-$1.750 & 0.309 & 9.267 & 9.547 & 8.803 & 0.487 & 0.580 & 0.374 & 0.246 & 0.065 & 4.630 & 2.470 & 0.295 & 0.357\\
1317846466364172800 & 244.858 & 29.713 & $-$1.690 & 0.332 & 11.321 & 11.462 & 11.046 & 0.493 & 0.567 & 0.362 & 0.212 & 0.091 & 4.795 & 2.598 & 0.252 & 0.365\\
5022411786734718208 & 26.249 & $-$30.059 & $-$1.750 & 0.377 & 11.303 & 11.435 & 11.050 & 0.464 & 0.553 & 0.341 & 0.153 & 0.081 & 4.834 & 3.223 & 0.245 & 0.367\\
5455339817249543040 & 158.011 & $-$30.177 & $-$1.485 & 0.330 & 11.779 & 11.895 & 11.500 & 0.433 & 0.427 & 0.307 & 0.158 & 0.145 & 4.803 & 3.244 & 0.267 & 0.368\\
6637515993378615168 & 279.419 & $-$57.461 & $-$1.098 & 0.323 & 14.042 & 14.201 & 13.748 & 0.421 & 0.509 & 0.289 & 0.098 & 0.088 & 5.164 & 4.043 & 0.257 & 0.370\\
4947090013255935616 & 39.274 & $-$42.963 & $-$1.480 & 0.311 & 8.937 & 9.066 & 8.668 & 0.465 & 0.562 & 0.340 & 0.163 & 0.073 & 4.590 & 2.782 & 0.258 & 0.373\\
2491648104802033920 & 29.468 & $-$5.534 & $-$1.291 & 0.301 & 11.966 & 12.108 & 11.703 & 0.369 & 0.450 & 0.279 & 0.103 & 0.071 & 4.763 & 3.543 & 0.250 & 0.376\\
4752258893572836352 & 42.892 & $-$47.801 & $-$1.110 & 0.311 & 12.282 & 12.431 & 12.010 & 0.410 & 0.494 & 0.306 & 0.149 & 0.089 & 4.661 & 3.386 & 0.255 & 0.383\\
6811546934337150464 & 325.526 & $-$25.475 & $-$1.230 & 0.317 & 11.550 & 11.688 & 11.255 & 0.362 & 0.482 & 0.272 & 0.091 & 0.112 & 4.196 & 3.817 & 0.273 & 0.384\\
3695934570706802944 & 190.604 & $-$0.201 & $-$1.419 & 0.361 & 14.499 & 14.645 & 14.225 & 0.406 & 0.480 & 0.286 & 0.132 & 0.042 & 4.865 & 3.325 & 0.262 & 0.392\\
2342755646078448000 & 12.371 & $-$27.387 & $-$1.282 & 0.355 & 13.283 & 13.430 & 13.014 & 0.432 & 0.491 & 0.289 & 0.088 & 0.061 & 5.128 & 5.452 & 0.260 & 0.393\\
4240897922656667648 & 300.509 & 2.357 & $-$1.847 & 0.345 & 13.415 & 13.654 & 13.006 & 0.568 & 0.660 & 0.424 & 0.256 & 0.096 & 4.575 & 2.375 & 0.298 & 0.394\\
5022005139231113088 & 25.610 & $-$30.460 & $-$1.831 & 0.377 & 12.251 & 12.394 & 11.981 & 0.425 & 0.517 & 0.322 & 0.093 & 0.066 & 4.883 & 3.652 & 0.262 & 0.395\\
2993559258622864000 & 93.232 & $-$14.668 & $-$1.384 & 0.320 & 11.813 & 12.049 & 11.407 & 0.382 & 0.464 & 0.297 & 0.136 & 0.062 & 4.583 & 3.163 & 0.301 & 0.399\\
4299563705588494848 & 301.653 & 9.498 & $-$1.255 & 0.330 & 11.738 & 11.929 & 11.364 & 0.390 & 0.385 & 0.286 & 0.115 & 0.105 & 4.973 & 4.468 & 0.302 & 0.399\\
2642479663953557888 & 358.534 & 0.964 & $-$1.195 & 0.306 & 10.498 & 10.646 & 10.203 & 0.379 & 0.412 & 0.278 & 0.142 & 0.075 & 4.804 & 4.024 & 0.277 & 0.402\\
1453674738379109760 & 209.392 & 29.858 & $-$1.230 & 0.329 & 11.294 & 11.416 & 10.996 & 0.366 & 0.521 & 0.306 & 0.076 & 0.074 & 4.824 & 4.454 & 0.292 & 0.404\\
5919795415987631104 & 262.383 & $-$55.805 & $-$1.697 & 0.327 & 13.017 & 13.212 & 12.672 & 0.481 & 0.564 & 0.342 & 0.234 & 0.150 & 5.055 & 3.366 & 0.286 & 0.404\\
5638021028504915840 & 140.252 & $-$26.741 & $-$1.372 & 0.356 & 13.245 & 13.423 & 12.918 & 0.445 & 0.533 & 0.301 & 0.084 & 0.091 & 5.004 & 4.263 & 0.286 & 0.410\\
3471695908730784896 & 186.820 & $-$29.084 & $-$1.867 & 0.370 & 13.686 & 13.860 & 13.374 & 0.508 & 0.611 & 0.403 & 0.213 & 0.115 & 4.434 & 2.405 & 0.283 & 0.420\\
4735802984075899904 & 52.921 & $-$52.344 & $-$1.220 & 0.338 & 14.047 & 14.200 & 13.757 & 0.385 & 0.453 & 0.264 & 0.077 & 0.084 & 5.001 & 4.118 & 0.281 & 0.423\\
6147659507088529408 & 182.090 & $-$43.067 & $-$1.446 & 0.279 & 13.222 & 13.431 & 12.856 & 0.424 & 0.508 & 0.325 & 0.176 & 0.042 & 4.515 & 2.270 & 0.300 & 0.423\\
4692528057537147136 & 19.628 & $-$67.918 & $-$1.540 & 0.406 & 11.534 & 11.702 & 11.231 & 0.421 & 0.482 & 0.287 & 0.068 & 0.147 & 6.518 & 4.975 & 0.290 & 0.441\\
3022122470468515328 & 88.161 & $-$5.859 & $-$1.699 & 0.374 & 11.920 & 12.250 & 11.404 & 0.370 & 0.465 & 0.291 & 0.131 & 0.051 & 4.623 & 3.148 & 0.348 & 0.460\\
5834468537710464640 & 243.026 & $-$59.714 & $-$1.428 & 0.375 & 11.679 & 11.905 & 11.269 & 0.414 & 0.515 & 0.306 & 0.063 & 0.067 & 5.310 & 4.306 & 0.336 & 0.468\\
5381396526408876416 & 178.087 & $-$40.488 & $-$1.418 & 0.340 & 14.080 & 14.374 & 13.608 & 0.381 & 0.459 & 0.285 & 0.150 & 0.078 & 4.667 & 3.051 & 0.362 & 0.514\\
\hline
\end{tabular}
\end{adjustbox}
\end{table*}

\section{All-sky extinction maps using \texttt{int\_average\_bp} 
and \texttt{int\_average\_rp} magnitudes}\label{sec:appendix_b}

In this Appendix, we present all-sky maps in Galactic coordinates of the A$_{G}$ values, 
derived from the color excesses $E(G - G_{\mathrm{RP}})$ and $E(G_{\mathrm{BP}} - G_{\mathrm{RP}})$ using the empirical relations obtained in this work. We adopted here for the $G_{BP}$ and $G_{RP}$ mean magnitudes the \texttt{int\_average\_bp} and  \texttt{int\_average\_rp} values from the {\it Gaia} DR3 \texttt{vari\_rrlyrae} table.\\
The maps are shown separately for RRab stars, for RRc stars (where AmpG was adopted to derive intrinsic colors $(G - G_{\mathrm{RP}})_{0}$), and for the combined RRab+RRc sample. 
Individual negative estimates, arised from photometric scatter and calibration uncertainties in low-extinction regions, are not included only for visualization purposes.
Color excesses were converted into absolute extinctions adopting the extinction coefficients from eqs.~\ref{eq:egrp} and~\ref{eq:ebprp} (see Section~\ref{sec:ris}), and assuming $R_V = 3.1$.

\begin{figure*}
\centering
\includegraphics[width=9.5cm]{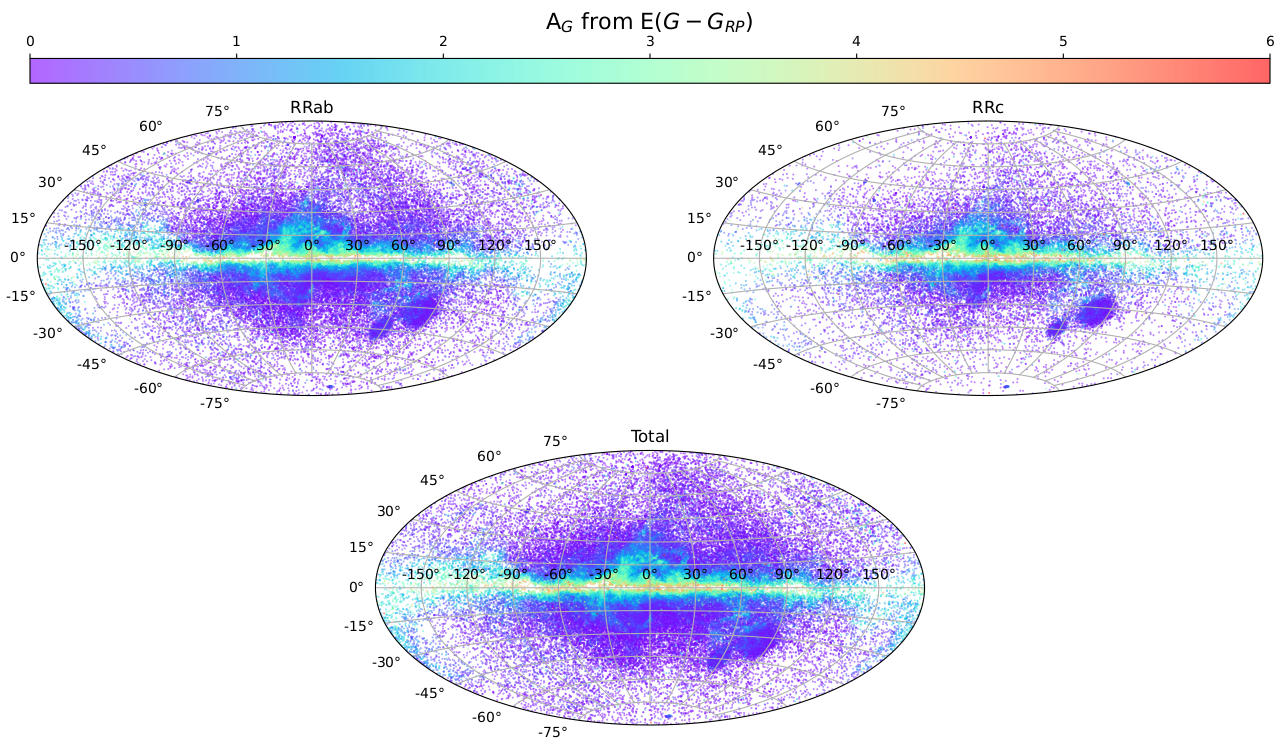}~\includegraphics[width=9.5cm]{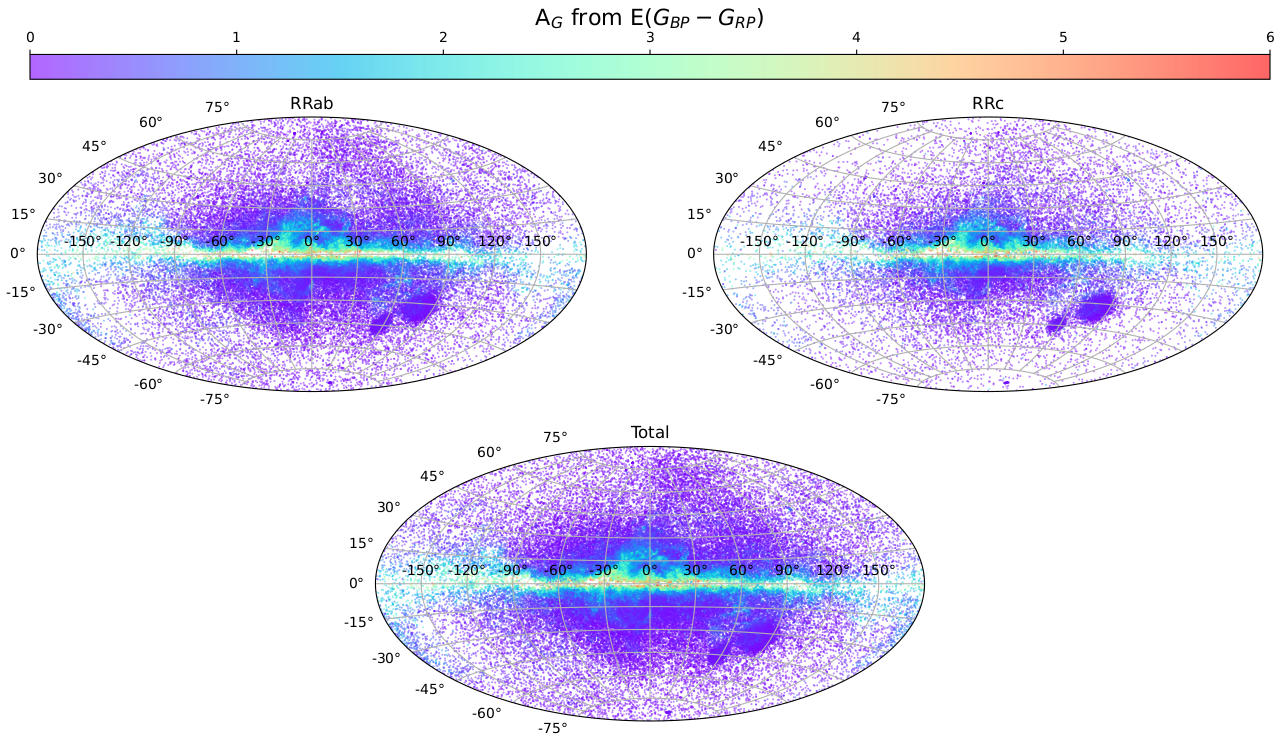}
\caption{ All-sky $A_G$ distribution in Galactic coordinates using \texttt{int\_average\_bp} and \texttt{int\_average\_rp} magnitudes from the \texttt{vari\_rrlyrae} table in {\it Gaia} DR3. {\it Left panels:} $A_G$ from $E(G-G_{\rm RP})$ for RRab (top left, $N_{\rm RRab} = 89917$, with $A_G \ge 0$ 82671), RRc (top right, $N_{\rm RRc} = 48672$, with $ A_G\ge 0$ 43974 using AmpG), and the combined RRab+RRc sample (bottom, $N_{\rm RRab+RRc} = 138589$). {\it Right panels:} $A_G$ from $E(G_{\rm BP}-G_{\rm RP})$, with $N_{\rm RRab} = 89915$ (83985 with $A_G \ge 0$), $N_{\rm RRc} = 48671$ (45907 with $ A_G\ge 0$), and $N_{\rm RRab+RRc} = 138586$. Individual negative $A_G$ values (amounting to a total of 11,944 sources for the $G-G_{\rm RP}$ map and 8,694 sources for the $G_{\rm BP}-G_{\rm RP}$ map) are not included in the plots for physical visualization; colors indicate extinction from 0 to 6 mag.}
      \label{fig:allsky_ag_bprpcu7}
\end{figure*} 

\section{Comparison of all-sky extinction maps using  \citet{SEF2011}}\label{sec:appendix_c}
\begin{figure*}
\centering
\includegraphics[width=9.4cm]{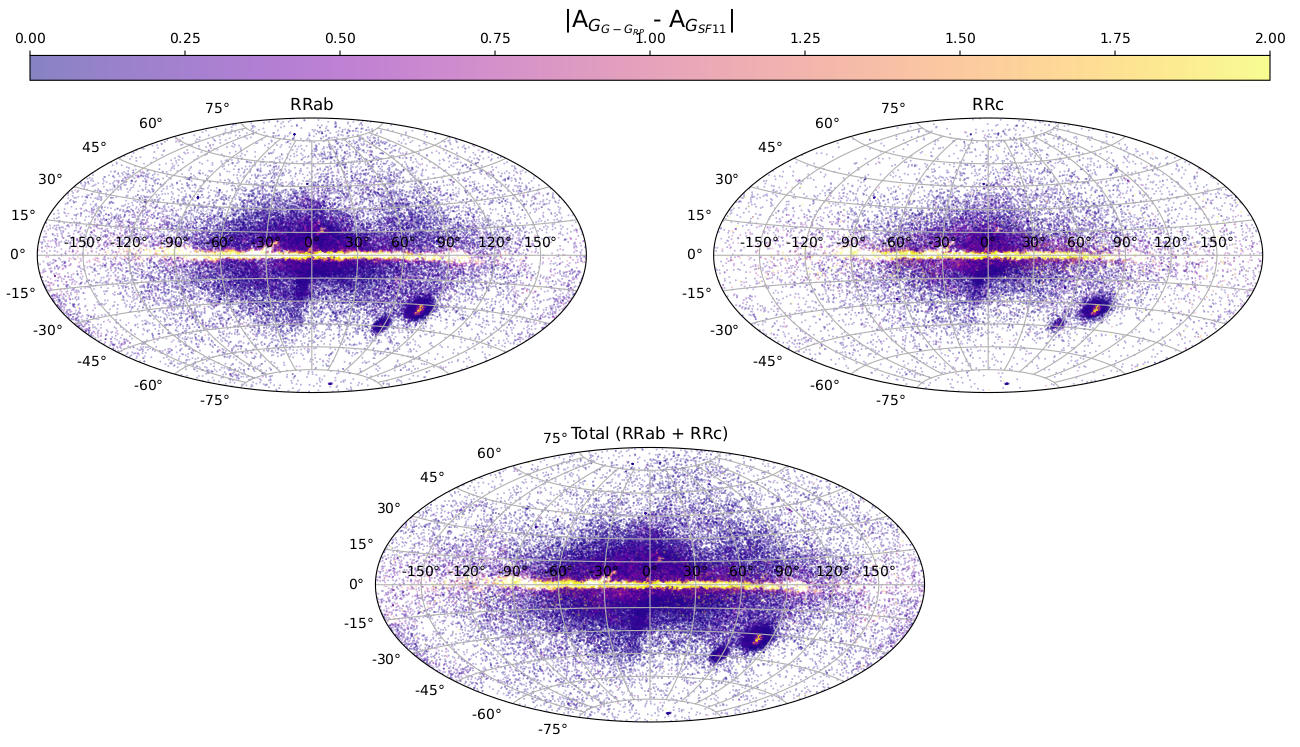}~\includegraphics[width=9.4cm]{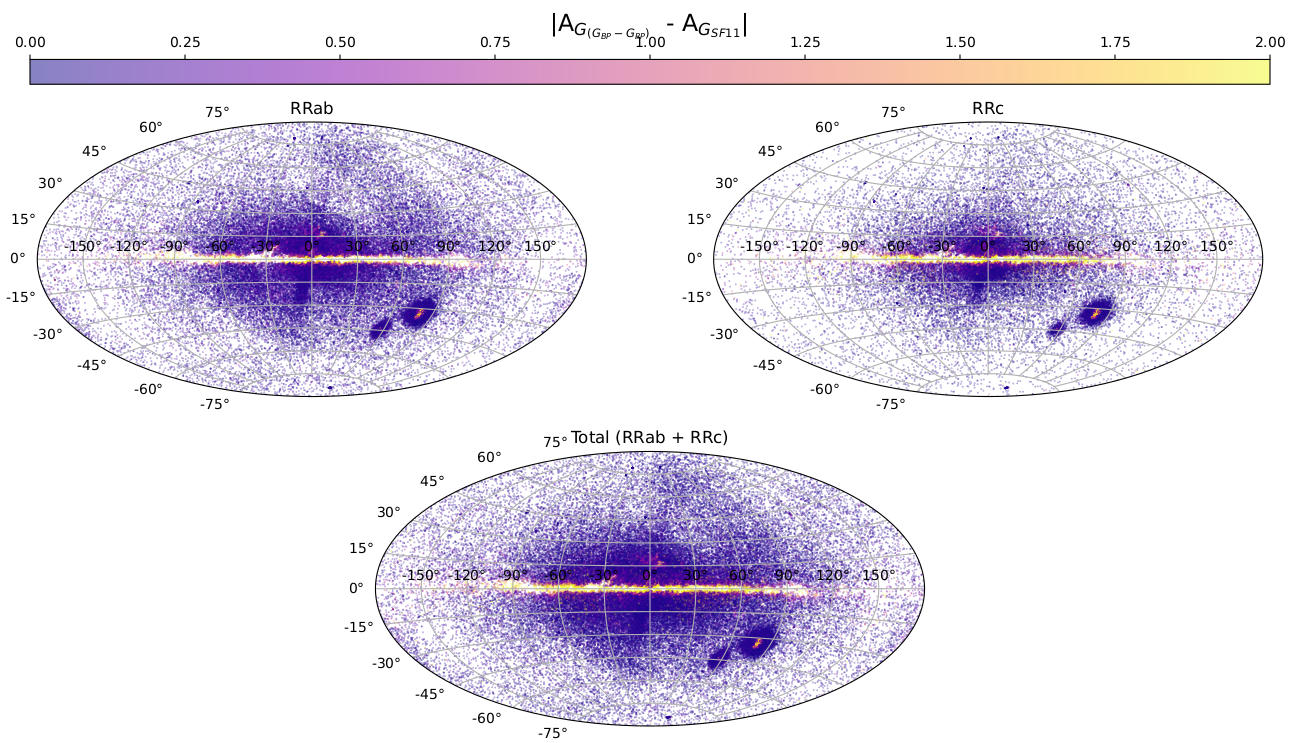}
      \caption{All-sky distribution in Galactic coordinates showing the absolute extinction discrepancy $|\Delta A_G| = |A_{G_\text{this work}} - A_{G_{SF11}}|$. The left set of panels displays $A_G$ derived from the $(G-G_{RP})_0$ relations, while the right set refers to the $(G_{BP}-G_{RP})_0$ calibrations. Within each set, results are presented for RRab stars (top left), RRc stars (top right), and the combined RRab+RRc sample (bottom center). The color scale is restricted to the 0--2~mag range to highlight the high degree of consistency between our stellar-derived values and the SF11 maps for the predominant portion of the sample. Darker regions represent areas of excellent agreement ($|\Delta A_G| \approx 0$), while the features emerging at low Galactic latitudes and toward the MCs indicate the transition toward the high-discrepancy regime.}
      \label{fig:com_allsky_ag_sef}
\end{figure*}

This appendix provides a visual comparison between the A$_{G}$ values derived in this work and the \citet[SF11]{SEF2011} 
maps. These figures support the discussion in Sect.~\ref{sec:ris} regarding the performance of our ML relations in high-reddening environments.
Fig.~\ref{fig:com_allsky_ag_sef} shows the absolute difference $|\Delta A_G| = |A_{G_\text{this work}} - A_{G_{SF11}}|$ across the entire sky for both RRab and RRc stars. Although the agreement is excellent over the majority of the celestial footprint, significant discrepancies emerge in the Galactic bulge and near the Galactic plane.

\begin{figure*}
\centering
\includegraphics[width=13cm]{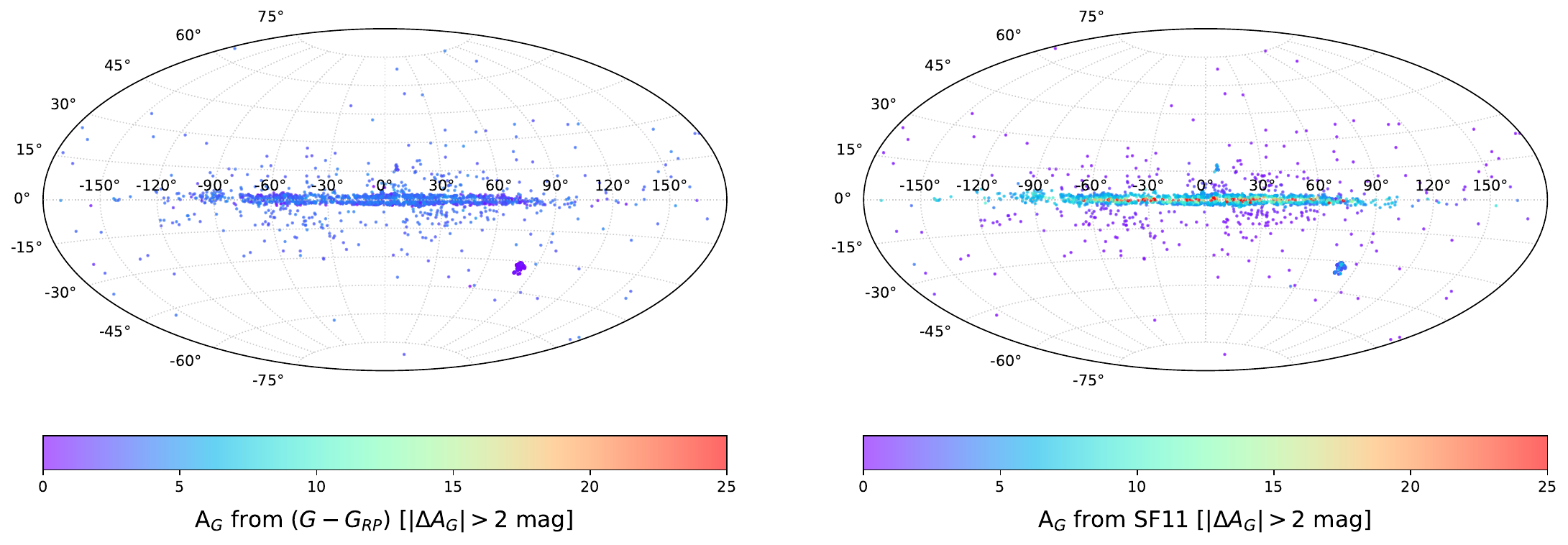}
\includegraphics[width=13cm]{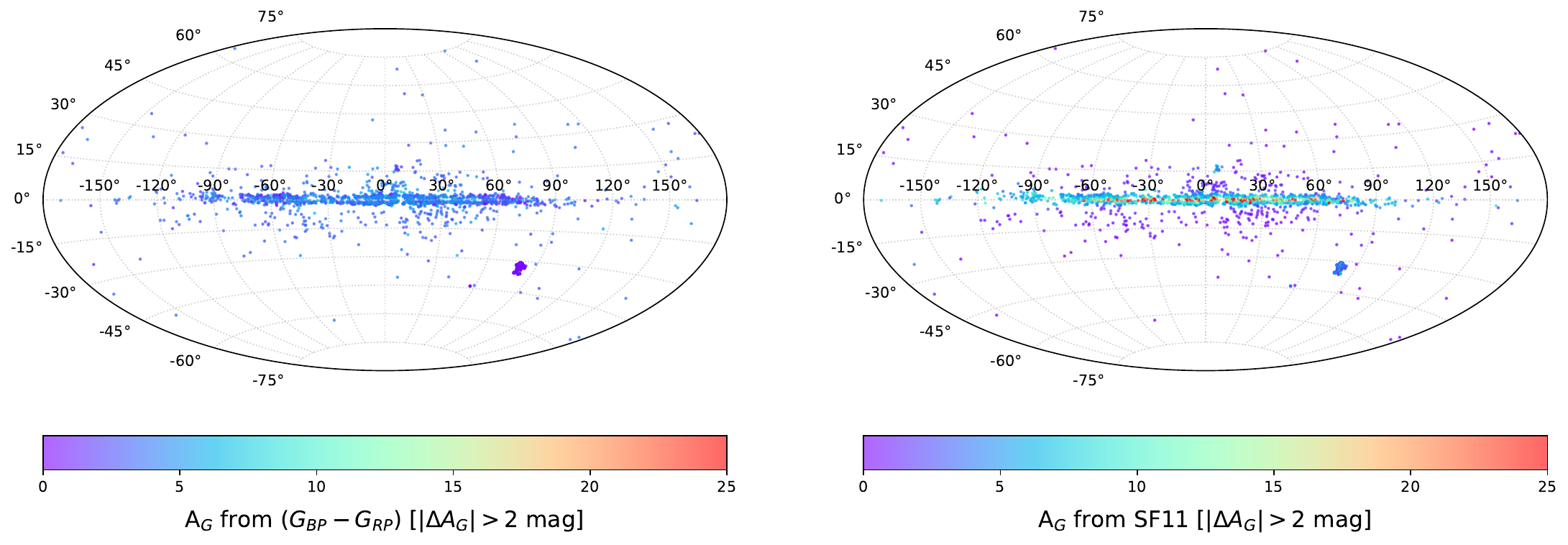}
      \caption{All-sky distribution in Galactic coordinates of sources with $|\Delta A_G| > 2$~mag. The left panels show the combined RRab+RRc sample color-coded by our $A_G$ values derived from the $(G-G_{RP})_0$ (upper, $N = 3295$) and $(G_{BP}-G_{RP})_0$ (lower, $N = 2868$) calibrations. The right panels display the same sources color-coded according to the SF11 extinction maps. The color scale is set to a maximum of 30~mag to accommodate the extreme $A_G$ values present in the SF11 dataset, which can exceed 25~mag in these regions.}
      \label{fig:com_allsky_ag_sef_2mag}
\end{figure*}

To further investigate these regions, in Fig.~\ref{fig:com_allsky_ag_sef_2mag} we plotted only the sources where $|\Delta A_G| >$2~mag. This subset allows for a direct visual comparison between our A$_{G}$ estimates (left panels) and those from SF11 (right panels), highlighting 
how both the $(G-G_{\rm RP})_0$ (3,295 sources) and $(G_{\rm BP}-G_{\rm RP})_0$ (2,868 sources) calibrations confine these larger discrepancies to the same highly localized, high-extinction regimes.

\end{appendix}
\end{document}